\documentclass[12pt,preprint]{aastex}

\usepackage{natbib}
\usepackage{rotating}
\usepackage{lscape}
\usepackage{geometry}
\begin{document}

\title{An Unbiased 1.3 mm Emission Line Survey of the Protoplanetary Disk Orbiting LkCa 15} 

\author{K. M. Punzi\altaffilmark{1}, P. Hily-Blant\altaffilmark{2}, J. H.\ Kastner\altaffilmark{1}, G.G. Sacco\altaffilmark{3}, T. Forveille\altaffilmark{2}}

\altaffiltext{1}{Center for Imaging Science, School of Physics \& Astronomy, and 
Laboratory for Multiwavelength Astrophysics, Rochester Institute of
Technology, 74 Lomb Memorial Drive, Rochester NY 14623 USA}
\altaffiltext{2}{UJF - Grenoble 1/CNRS-INSU, Institut de Plan\'{e}tologie et d'Astrophysique de Grenoble (IPAG) UMR 5274, 38041, Grenoble, France}
\altaffiltext{3}{INAF - Osservatorio Astrofisico di Arcetri, Largo E. Fermi 5, 50125, Firenze, Italy}

\begin{abstract}
The outer ($>$30 AU) regions of the dusty circumstellar disk orbiting the $\sim$2--5 Myr-old, actively accreting solar analog LkCa 15 are known to be chemically rich, and the inner disk may host a young protoplanet within its central cavity.
To obtain a complete census of the brightest molecular line emission emanating from the LkCa 15 disk over the 210-270 GHz (1.4 - 1.1 mm) range, we have conducted an unbiased radio spectroscopic survey with the Institute de Radioastronomie Millim\'{e}trique (IRAM) 30 meter telescope.
The survey demonstrates that, in this spectral region, the most readily detectable lines are those of CO and its isotopologues ${\rm ^{13}CO}$ and ${\rm C^{18}O}$, as well as ${\rm HCO^+}$, HCN, CN, ${\rm C_2H}$, CS, and ${\rm H_2CO}$.
All of these species had been previously detected in the LkCa 15 disk; however, the present survey includes the first complete coverage of the CN (2--1) and ${\rm C_2H}$ (3--2) hyperfine complexes.
Modeling of these emission complexes indicates that the CN and ${\rm C_2H}$ either reside in the coldest regions of the disk or are subthermally excited, and that their abundances are enhanced relative to molecular clouds and young stellar object environments.
These results highlight the value of unbiased single-dish line surveys in guiding future high resolution interferometric imaging of disks.
\end{abstract}

\section{Introduction}

To understand the origins of our own solar system, it is necessary to investigate the protoplanetary disks of young, solar-mass stars (T Tauri stars) that represent early solar analogs.
The LkCa 15 star/disk system, which has been extensively studied at X-ray, optical, infrared, and (sub)millimeter wavelengths, represents a prime example of such an early solar analog \citep{2004A&A...425..955T,2010ApJ...720..480O,2011ApJ...742L...5A,2012ApJ...747..136I,2013ApJ...765....3S,2014A&A...566A..51T}.
The LkCa 15 system is associated with the Taurus-Auriga star-forming cloud complex, at a distance of about 140 pc \citep{1998A&A...330..145V}, and is an archetypal ``transition disk'' system (see, e.g., \citealt*{2011ARA&A..49...67W}).
The K5 ($L_{bol} = 0.74 L_{\odot}$) central star is estimated to have an age of 2-5 Myr and a mass of M = 1.0 M$_{\odot}$ (\citealt{2000ApJ...545.1034S}; \citealt*{1995ApJS..101..117K}).
The LkCa 15 star/disk system is characterized by a mass accretion rate of about $3 \times 10^{-9}$ M$_{\odot}$ yr$^{-1}$ \citep{1998ApJ...495..385H}, a partially dust-depleted inner region of $\sim$50 AU in radius \citep{2011ApJ...742L...5A,2007ApJ...670L.135E,2006A&A...460L..43P}, and an infrared excess over the stellar photosphere that can be explained by the presence of hot dust within a few AUs from the central star \citep{2007ApJ...670L.135E,2010A&A...512A..11M}.
The disk, which displays detectable molecular gas in Keplerian rotation out to $\sim$900 AU \citep{2007A&A...467..163P}, is viewed at intermediate inclination ($i \sim 52^\circ$; \citealt{2011A&A...535A.104D}).

The $\sim$50 AU central cavity constitutes one of the largest inner submm ``holes'' observed among transition disks \citep{2011ApJ...742L...5A}.
The discrepancy in the outer disk radius as inferred from interferometry of emission from dust and gas \citep[150 AU vs.\ 900 AU, respectively;][]{2007A&A...467..163P,2012ApJ...747..136I} also suggests the density of mm-size dust grains drops precipitously in the outer disk, beyond $\sim$150 AU.
This radially stratified dust disk structure may reflect the effects of recent or ongoing planet formation.
Indeed, \citet*{2012ApJ...745....5K} reported the potential discovery of a 6 M$_{J}$, $\sim$ 1 Myr old protoplanet well within the submm continuum cavity (i.e., at a distance of 16 AU from the central T Tauri star).
It is unclear how a cavity as large as that observed within the LkCa 15 disk may be formed, but it is believed that it must be due to more than just dynamical interactions with a single planet, suggesting the presence of multiple planets and/or the need for additional clearing mechanisms (e.g., grain growth, photoevaporation; \citealt{2011ApJ...729...47Z}; \citealt*{2011ApJ...738..131D}).

Given its status as the first accreting T Tauri star that has been identified as potentially harboring a young protoplanet, the LkCa 15 system serves as a unique laboratory for constraining physical conditions within a planet-forming disk --- one in which planet formation is likely at an advanced stage.
Indeed, among protoplanetary disks, the LkCa 15 disk is especially chemically rich; it has previously been detected in several molecular transitions \citep{2004A&A...425..955T,2007A&A...467..163P,2010ApJ...720..480O}.
Furthermore, \citet*{2013ApJ...765....3S} report the detection of LkCa 15 as a bright X-ray source with Chandra (intrinsic X-ray luminosity of log L$_{x}$ $\approx$ 30.4 ergs s$^{-1}$).
As discussed in \citet{2014ApJ...793...55K} and references therein, such strong X-ray emission (as well as extreme untraviolet emission) from the central star likely has important consequences for disk heating, chemistry, accretion, and mass-loss --- and, hence, for planet formation.

We have demonstrated that unbiased line surveys of evolved, chemically rich, protoplanetary disks, such as that orbiting LkCa 15, are necessary to establish the full inventory of readily detectable molecular species in disks \citep{2014ApJ...793...55K}.
Such observations are important both in providing constraints on the latest generations of disk models that incorporate stellar irradiation and gas-grain reactions, and in serving as guidance for subsequent interferometric imaging of molecular emission from protoplanetary disks \citep{Kastner2015}.
With this as motivation, we have carried out such an unbiased radio molecular line survey of the LkCa 15 disk with the 30 m IRAM radio telescope.

\section{Observations and Data Reduction}

\begin{figure}[tbp]
  \centering
  \includegraphics[width=0.9\hsize]{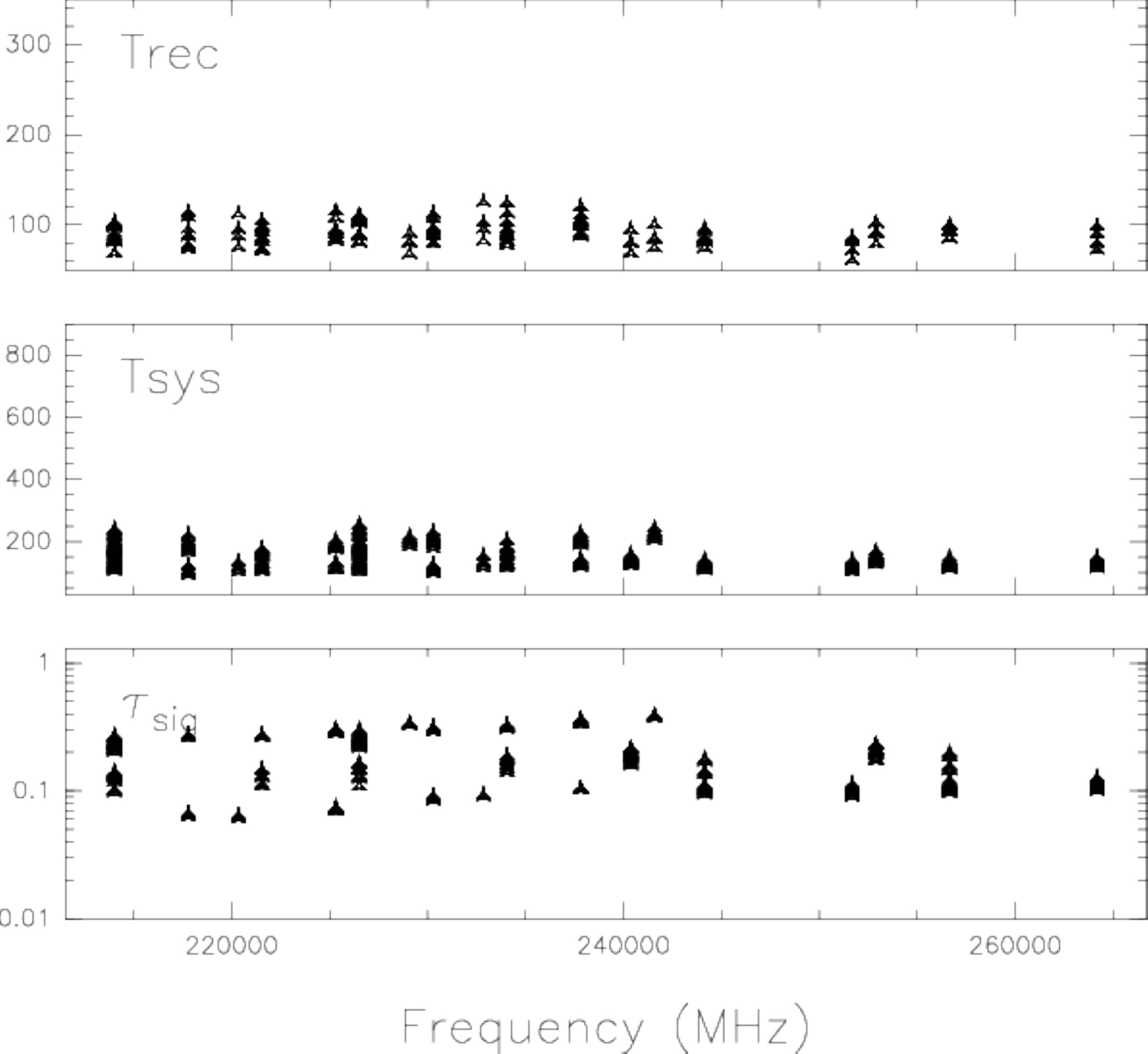}
  \caption{Receiver temperatures (top), system temperatures (middle), and signal-band opacities (bottom) as functions of frequency during the LkCa 15 spectral survey.}
  \label{fig:cal}
\end{figure}

\begin{figure}[tbp]
  \centering
  \includegraphics[width=0.9\hsize]{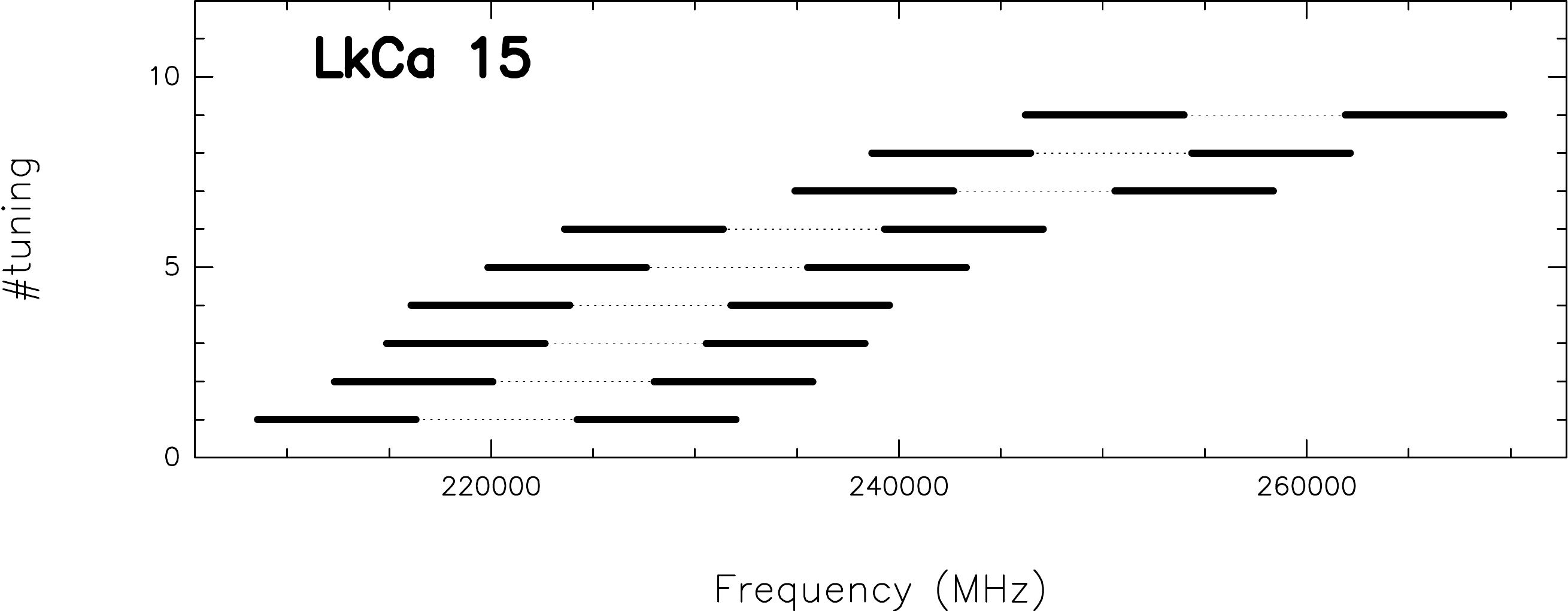}
  \caption{IRAM EMIR spectral coverage for the LkCa 15 spectral survey.}
  \label{fig:setup}
\end{figure}

The line survey of LkCa 15 reported here was carried out over the frequency range $\sim210-270$ GHz with the IRAM 30 m telescope and its EMIR receivers in March 2012 during good weather conditions ($\sim2.0$ mm of precipitable water vapor).
The EMIR receiver temperatures were fairly homogeneous over the frequency range surveyed (see Figure~\ref{fig:cal}), with values in the range 80--100~K.
The standard three-phase calibration procedure at the IRAM 30 m was used to obtain the zenith opacity every 10--15~min, with measured values between 0.1 and 0.3.
The EMIR receivers were used along with the Fourier Transform Spectrometers in their 200~kHz (0.25 km s$^{-1}$) spectral resolution mode, providing $\sim16$~GHz instantaneous bandwidth that was split into two chunks of $\sim8$ GHz each (see Figure~\ref{fig:setup}).
To ensure we obtained flat baselines and optimal atmospheric subtraction, the observations were performed in the wobbler switch mode, with a 70'' throw.
The pointing was checked approximately every 1--1.5~hours on a nearby continuum source.
The focus was checked on a strong source every 2--2.5~hours.
The frequency setups were chosen to ensure a factor two redundancy, except at the edges.
However, bad weather conditions reduced the observing time, and the final frequency coverage was reviewed accordingly (see Figure~\ref{fig:setup}).
Consequently, 9 setups have been observed so as to obtain a complete coverage from 208.546 to 269.666~GHz.
We note that having a large instantaneous bandwidth not only minimizes inter-calibration issues between lines, but also minimizes pointing issues.

Data reduction was performed with the CLASS software within the GILDAS package\footnote{\footnotesize http://www.iram.fr/IRAMFR/GILDAS}.
Residual bandpass effects were subtracted by applying a first order polynomial to each spectrum.
We find the noise to be almost uniform (see Figure~\ref{fig:cal}), with an average noise level of 7~mK (antenna temperature scale) in 0.27 km s$^{-1}$ channels across the full bandwidth.

The output of the foregoing data acquisition and reduction sequences for LkCa 15 is a single spectrum that covers the full $\sim$ 210-270 GHz frequency range.
The results for line intensities and derived quantities described in Section~\ref{sec:resultsanalysis} were obtained from these spectra, adopting a beam full width at half maximum (FWHM) of 10.7''.
The line intensities are quoted in main beam temperature scale, using $T_{mb} = \frac{F_{eff}}{B_{eff}}T_{A}^{*}$, where the appropriate values of $F_{eff}$ (forward efficiency) and $B_{eff}$ (main beam efficiency) were found by linear interpolation of the values that are provided in the IRAM 30 m documentation.
These line intensities can be converted to Jy km s$^{-1}$ using $3.906 \frac{F_{eff}}{A_{eff}}$ Jy/K for $S/T_{A}^{*}$, the point source sensitivity, with $F_{eff} =0.88-0.94$ and $A_{eff} = 0.41-0.49$ (aperture efficiency), where these represent the values for the IRAM 30 m telescope in the 1.3 mm window\footnote{\footnotesize See http://www.iram.es/IRAMES/mainWiki/Iram30mEfficiencies.}.

\section{Results and Analysis}
\label{sec:resultsanalysis}

\subsection{Molecular Line Inventory}

\begin{figure}[tbp]
   \caption{Spectra of all molecular transitions that have been studied in our IRAM 30m line survey of LkCa 15. Ordinate is velocity with respect to the local standard of rest (in km s$^{-1}$) and abscissa is line flux (in K); for clarity, spectral regions (other than ${\rm ^{12}CO}$) have been shifted upwards, and lines have been rescaled.}
  \label{fig:fullspecta}
  \centering
     \includegraphics[scale=4.0, angle=270, width=1.0\textwidth]{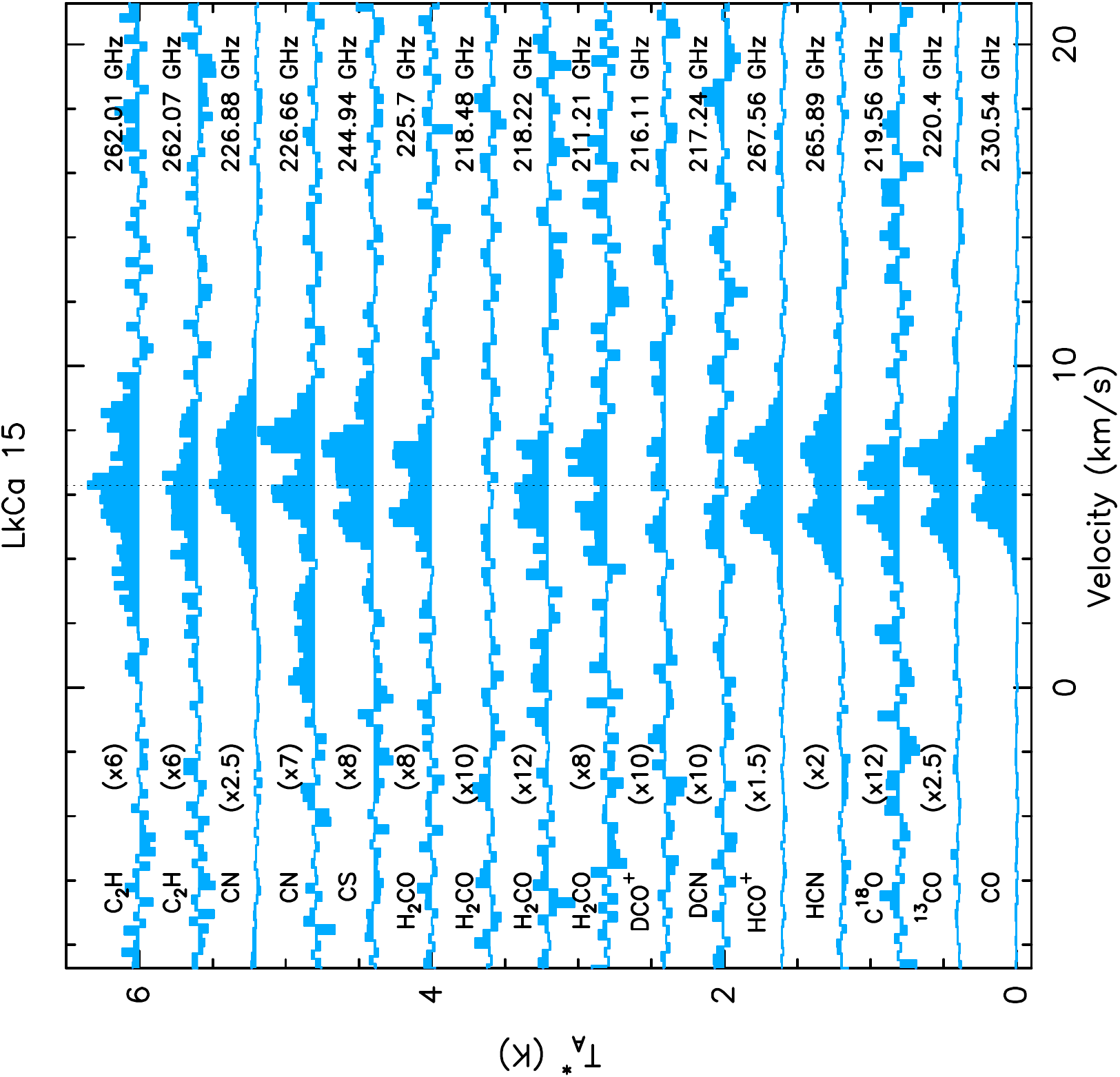}
\end{figure}

We used the spectral line identification methods available in the WEEDS extension of the GILDAS\footnote{\footnotesize See \citealt{2011A&A...526A..47M}.} software tools, along with visual inspection, to compile a list of molecular transitions that are readily detectable and measurable in the 210-270 GHz IRAM spectrum of LkCa 15.
A summary of these molecular transitions is presented in Table~\ref{tbl:MyDetections} and Table~\ref{tbl:CNC2HDetections}.
The spectral regions covering all the lines we detected are displayed in Figure~\ref{fig:fullspecta}.
A listing of all molecular transitions previously detected and searched for in LkCa 15 is presented in Table~\ref{tbl:MolecularSpeciesALLPub}.
An inspection of Table~\ref{tbl:MyDetections} shows that, in the spectral region surveyed, the most readily detectable lines are those of CO and its isotopologues ${\rm ^{13}CO}$ and ${\rm C^{18}O}$, as well as ${\rm HCO^+}$, HCN, CN, ${\rm C_2H}$, CS,  and ${\rm H_2CO}$.
All of these species had been previously detected in the LkCa 15 disk.
In most cases, however, the detections were widely spaced in time and/or made with different telescope/instrument combinations, potentially resulting in significant calibration uncertainties where, e.g., measurement of line ratios is concerned.

Absent such uncertainties we can, e.g., exploit our near-simultaneous coverage of CO and its isotopologues ${\rm ^{13}CO}$ and ${\rm C^{18}O}$ to estimate the optical depths of the observed transitions of CO and ${\rm ^{13}CO}$ (\S~\ref{sec:OpticalDepthApprox}).
In addition, the survey includes first measurements of the full suite of hyperfine transitions of CN $N = 2 \rightarrow 1$ and ${\rm C_2H}$ $N = 3 \rightarrow 2$, whose analysis yields estimates of optical depth and determinations of excitation temperature for these molecules (\S~\ref{sec:CNC2H}).
We note that the $J=5 \rightarrow 4$ transition of CS is detected here for the first time in the LkCa 15 disk.
We also detect multiple transitions of ${\rm H_2CO}$, one of which (the $J=3_{13}\rightarrow 2_{12}$ line at 211.211 GHz) had not been previously detected in this disk.

\subsection{Line Intensity Measurements}
\label{sec:LineIntensityMeasurements}

Because the LkCa 15 disk is viewed at intermediate inclination ($i \sim 52^\circ$; \citealt{2011A&A...535A.104D}), it displays molecular line profiles that are double-peaked.
Thus, we fit the observed profile with a Keplerian disk model described in Section 2 of \citet*{1993ApJ...402..280B} and used by \citet{2008A&A...492..469K,2010ApJ...723L.248K,2014ApJ...793...55K} to analyze molecular emission of protoplanetary disks observed with a single-dish radio telescope.
The profile parameters are: the peak line intensity ($T_{mb}$), the Keplerian velocity at the outer edge of the disk ($v_{d}$, equivalent to half of the peak-to-peak velocity width), the disk radial temperature profile power law component ($q$, where $T(r) \varpropto r^{-q}$), and the outer disk density cutoff ($p_{d}$).
When we have obtained a good fit to the observed line profile, as determined by the reduced $\chi^{2}$ of the fit, we can then integrate under the model line profile and obtain the total line intensity ($I$).
We stress that the purpose of applying such a (parametric) model is to obtain line intensity measurements, and not necessarily to infer disk structure; such inferences are, in any case, inherently difficult given only single-dish observations.

\begin{figure}[tbp]
 \caption{IRAM 30m spectra of ${\rm ^{12}CO}$ ($J=2\rightarrow 1$) (top), ${\rm ^{13}CO}$ ($J=2\rightarrow 1$) (middle), and ${\rm C^{18}O}$ ($J=2\rightarrow 1$) (bottom) detected in LkCa 15 (blue histogram) overlaid with parametric Keplerian model line profiles (see \S~\ref{sec:LineIntensityMeasurements}). Ordinate is velocity with respect to the local standard of rest (in km s$^{-1}$, with frequencies in MHz at the top of the panel) and abscissa is line flux (in K).}
 \label{fig:CO}
  \centering
    \includegraphics[scale=0.45, angle=270]{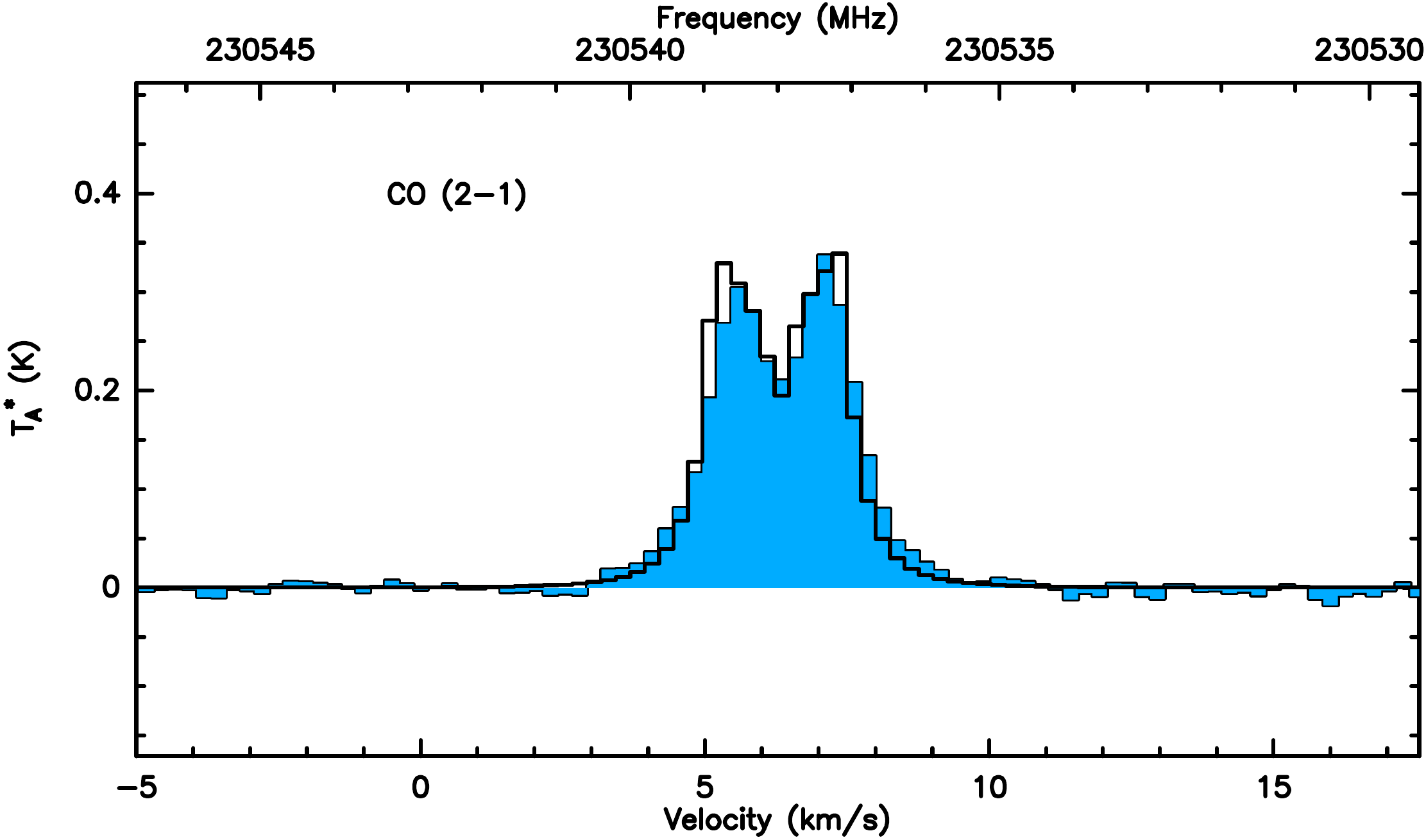}
    \includegraphics[scale=0.45, angle=270]{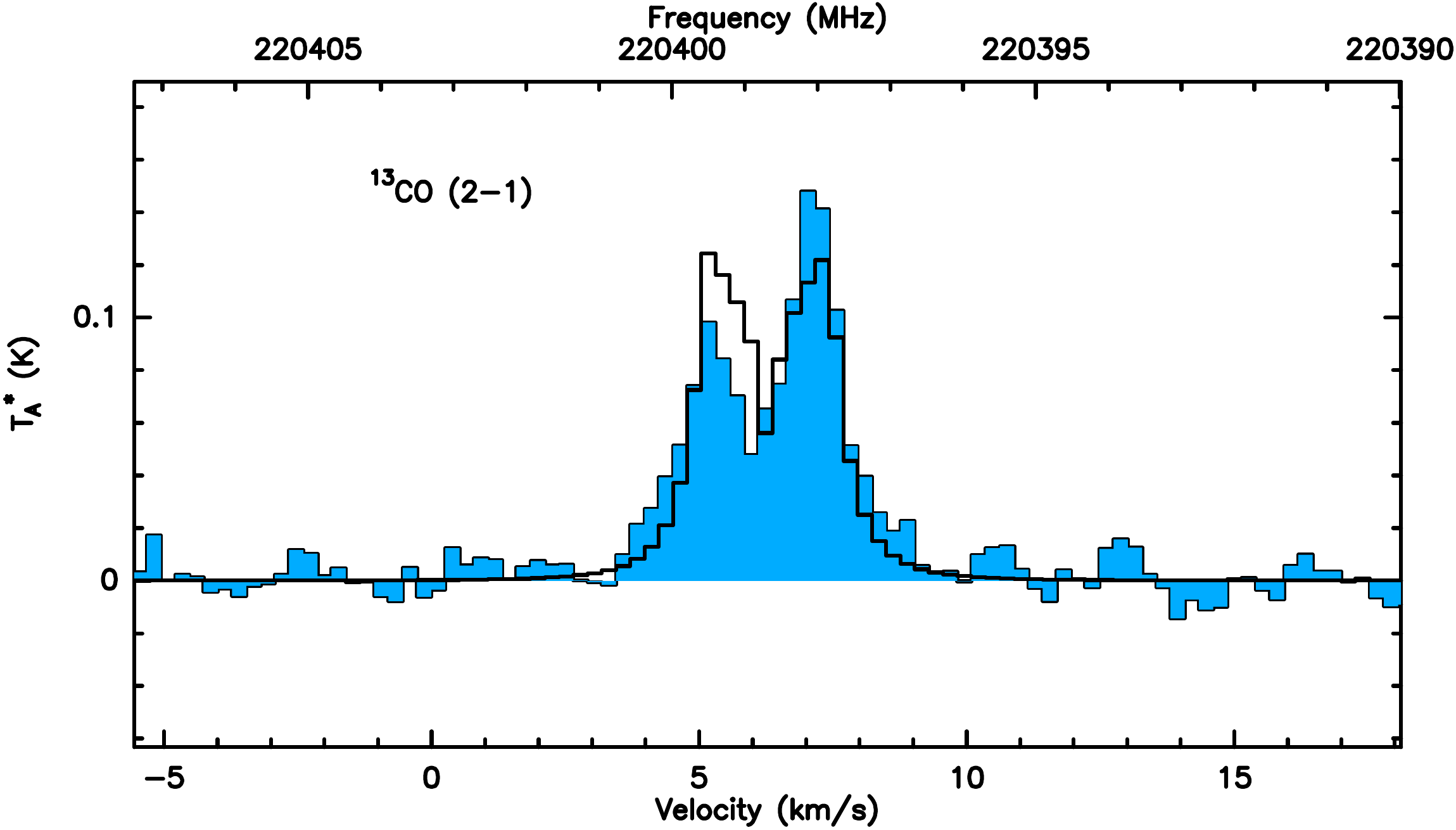}
    \includegraphics[scale=0.45, angle=270]{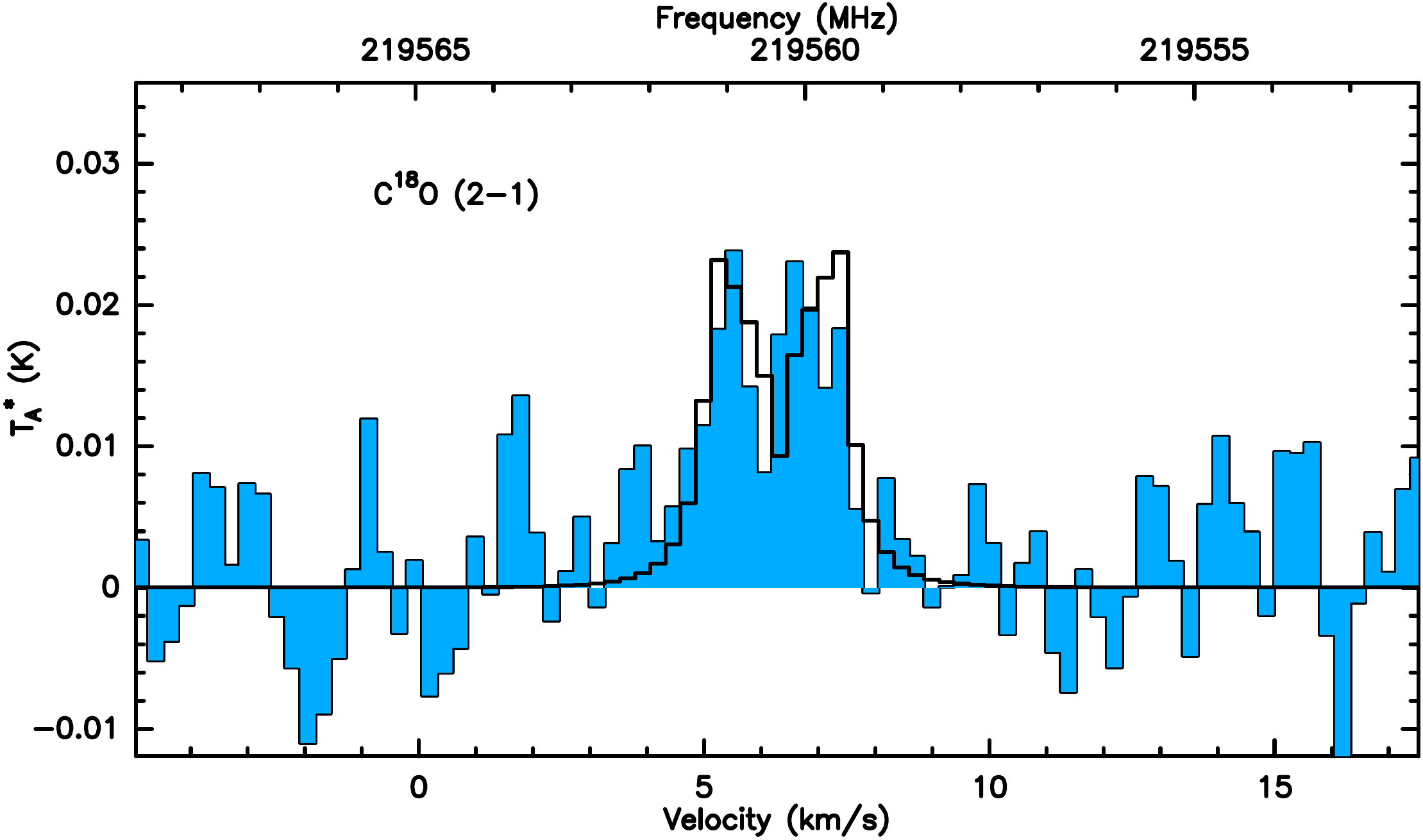}
\end{figure}

\begin{figure}[tbp]
  \caption{IRAM 30m spectra of ${\rm HCO^+}$ ($J=3\rightarrow 2$) (top),  HCN ($J=3\rightarrow 2$) (middle), and CS ($J=5\rightarrow 4$) (bottom) detected in LkCa 15 (blue histogram) overlaid with parametric Keplerian model line profiles (see \S~\ref{sec:LineIntensityMeasurements}). Ordinate is velocity with respect to the local standard of rest (in km s$^{-1}$, with frequencies in MHz at the top of the panel) and abscissa is line flux (in K).}
 \label{fig:HCO+HCNCS}
  \centering
      \includegraphics[scale=0.45, angle=270]{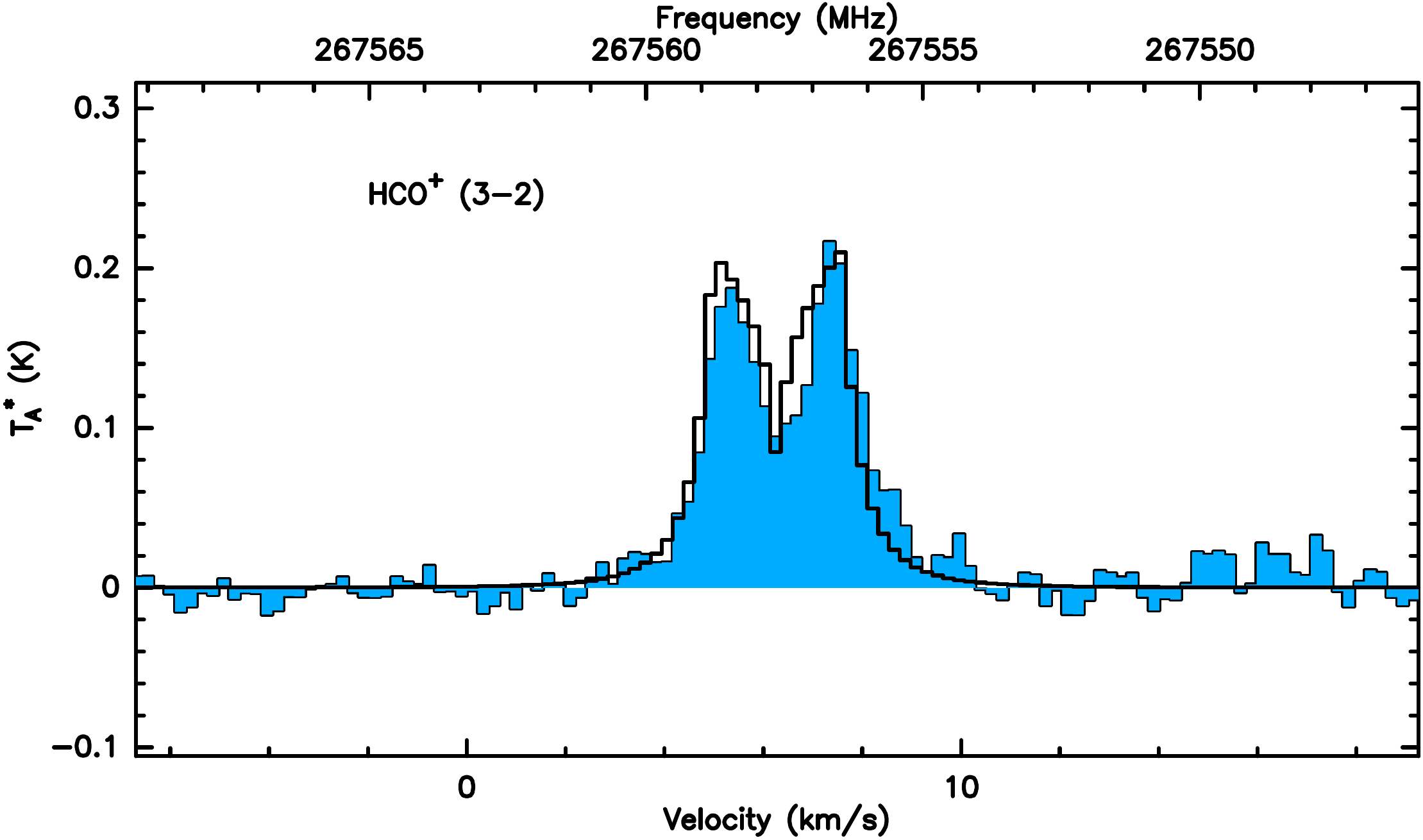}
      \includegraphics[scale=0.45, angle=270]{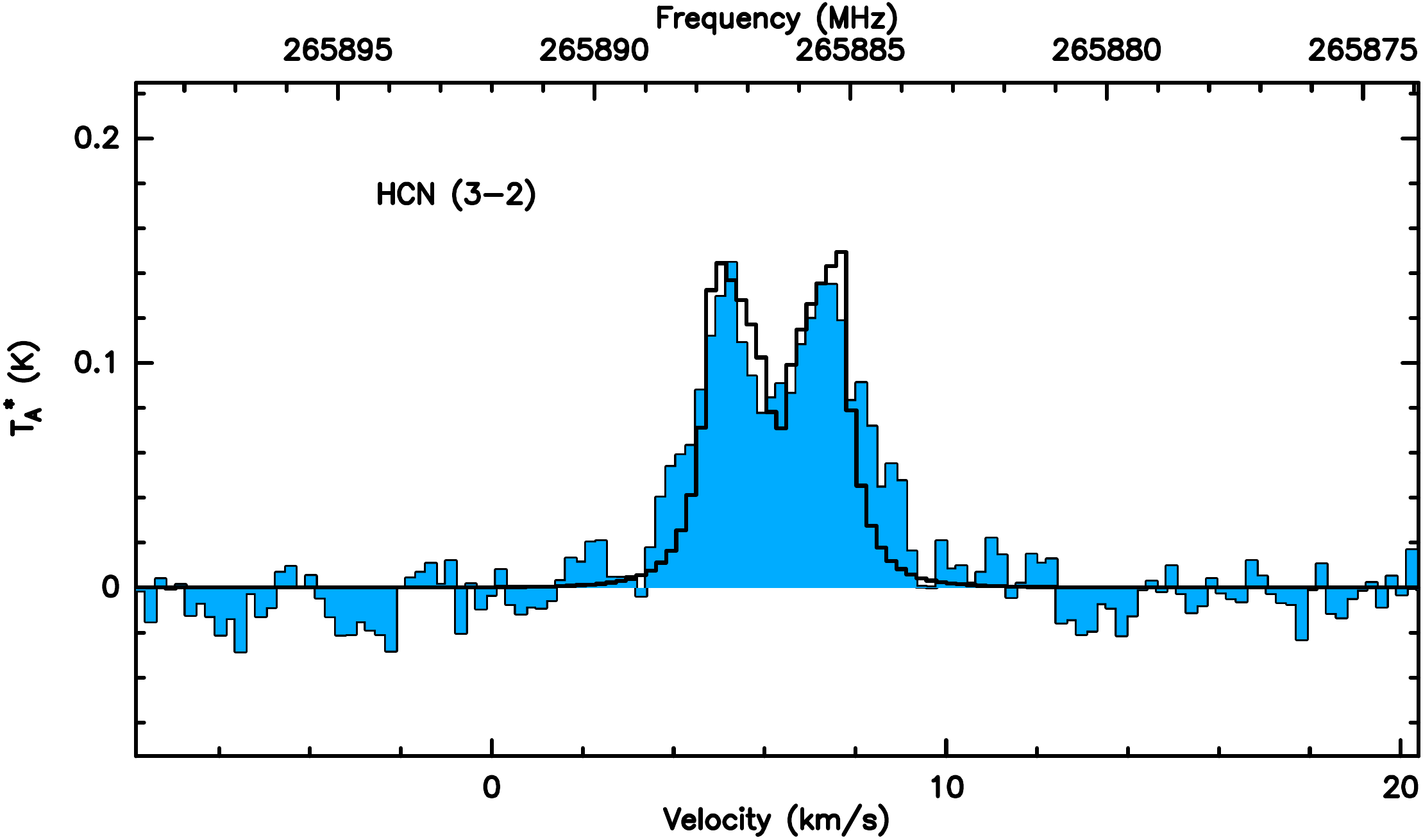}
      \includegraphics[scale=0.45, angle=270]{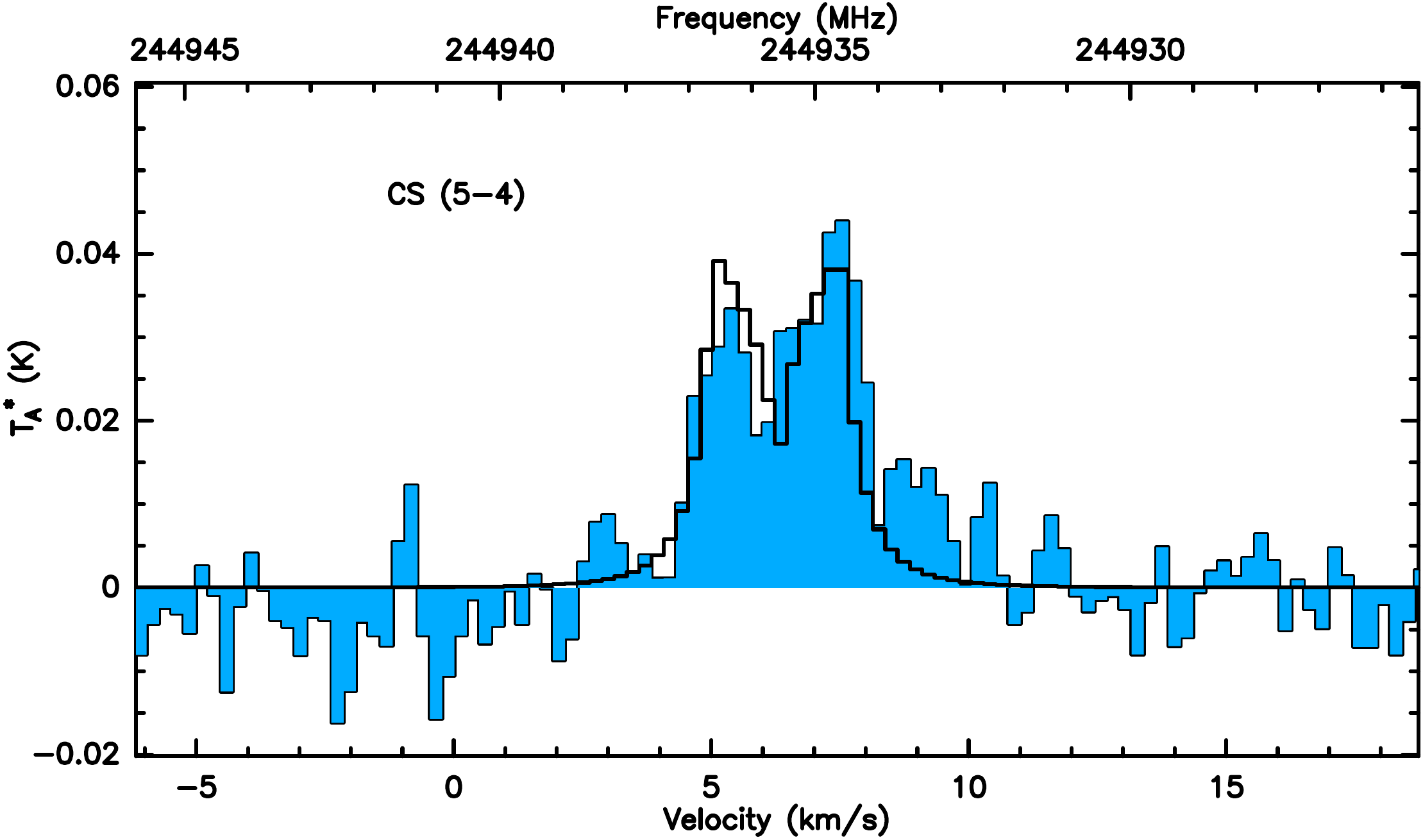}
\end{figure}

\begin{figure}[tbp]
  \caption{IRAM 30m spectra of ${\rm H_2CO}$ ($J=3_{13}\rightarrow 2_{12}$) (top),  ${\rm H_2CO}$ ($J=3_{03}\rightarrow 2_{02}$) (middle), and ${\rm H_2CO}$ ($J=3_{12}\rightarrow 2_{11}$) (bottom) detected in LkCa 15 (blue histogram) overlaid with parametric Keplerian model line profiles (see \S~\ref{sec:LineIntensityMeasurements}). Ordinate is velocity with respect to the local standard of rest (in km s$^{-1}$, with frequencies in MHz at the top of the panel) and abscissa is line flux (in K).}
 \label{fig:H2CO}
  \centering
       \includegraphics[scale=0.4, angle=270]{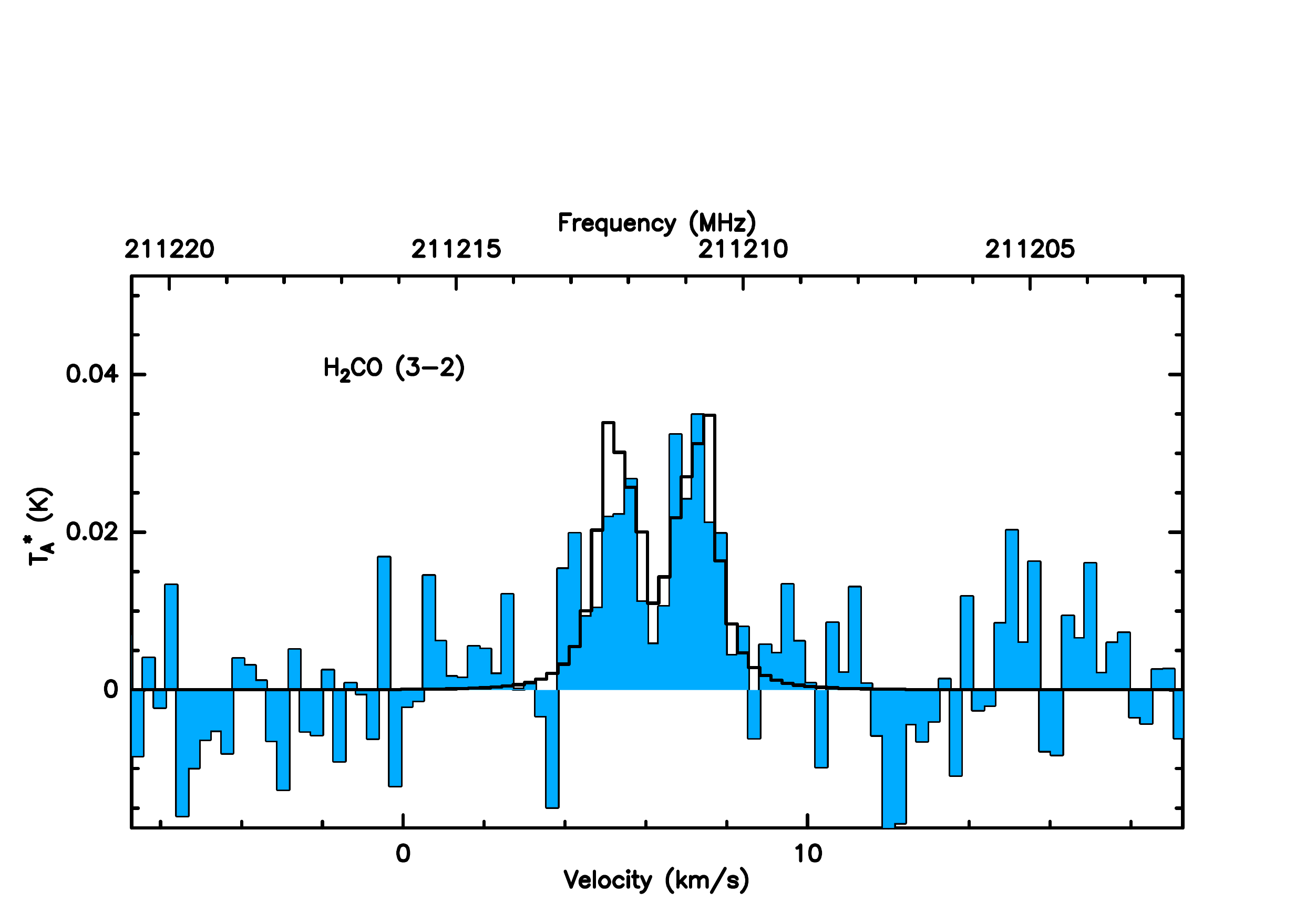}
       \includegraphics[scale=0.4, angle=270]{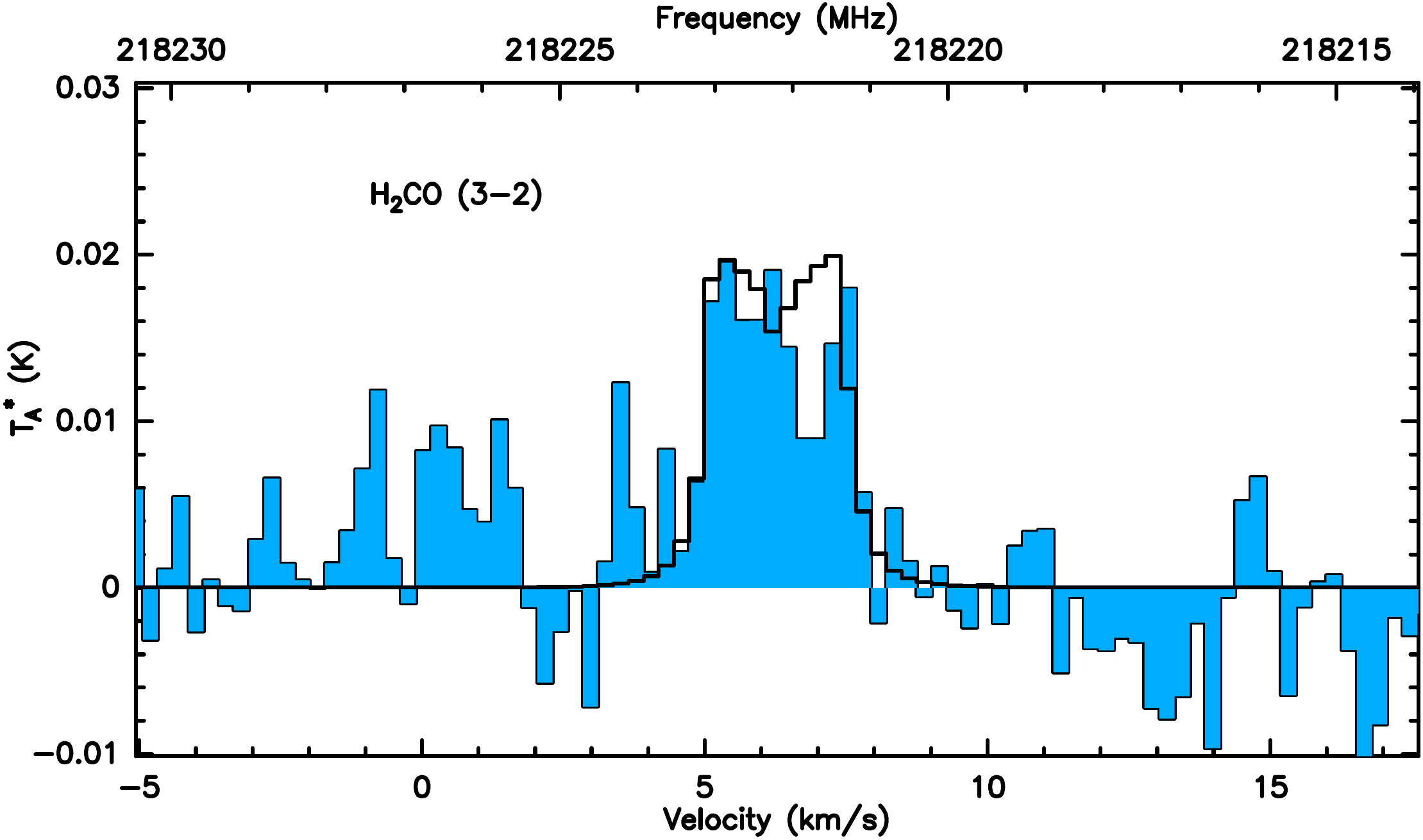}
       \includegraphics[scale=0.4, angle=270]{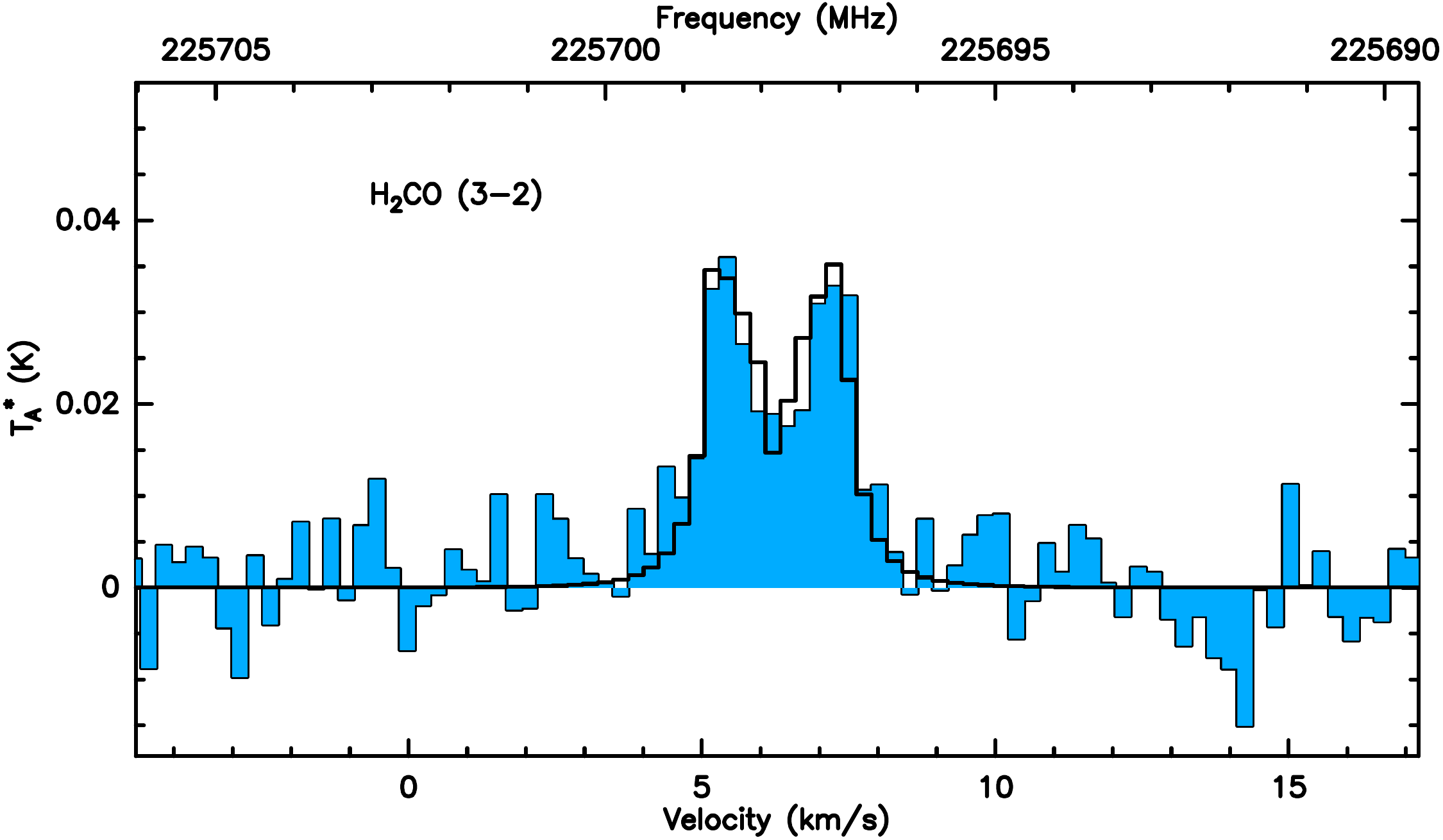}
\end{figure}

\begin{figure}[tbp]
  \caption{IRAM 30m smoothed spectra of DCN ($J=3\rightarrow 2$) (top) and ${\rm DCO^+}$ ($J=3\rightarrow 2$) (bottom) detected in LkCa 15 (blue histogram) (see \S~\ref{sec:LineIntensityMeasurements}). Ordinate is velocity with respect to the local standard of rest (in km s$^{-1}$, with frequencies in MHz at the top of the panel) and abscissa is line flux (in K).}
 \label{fig:tentativedetections}
  \centering
    \includegraphics[scale=0.45, angle=270]{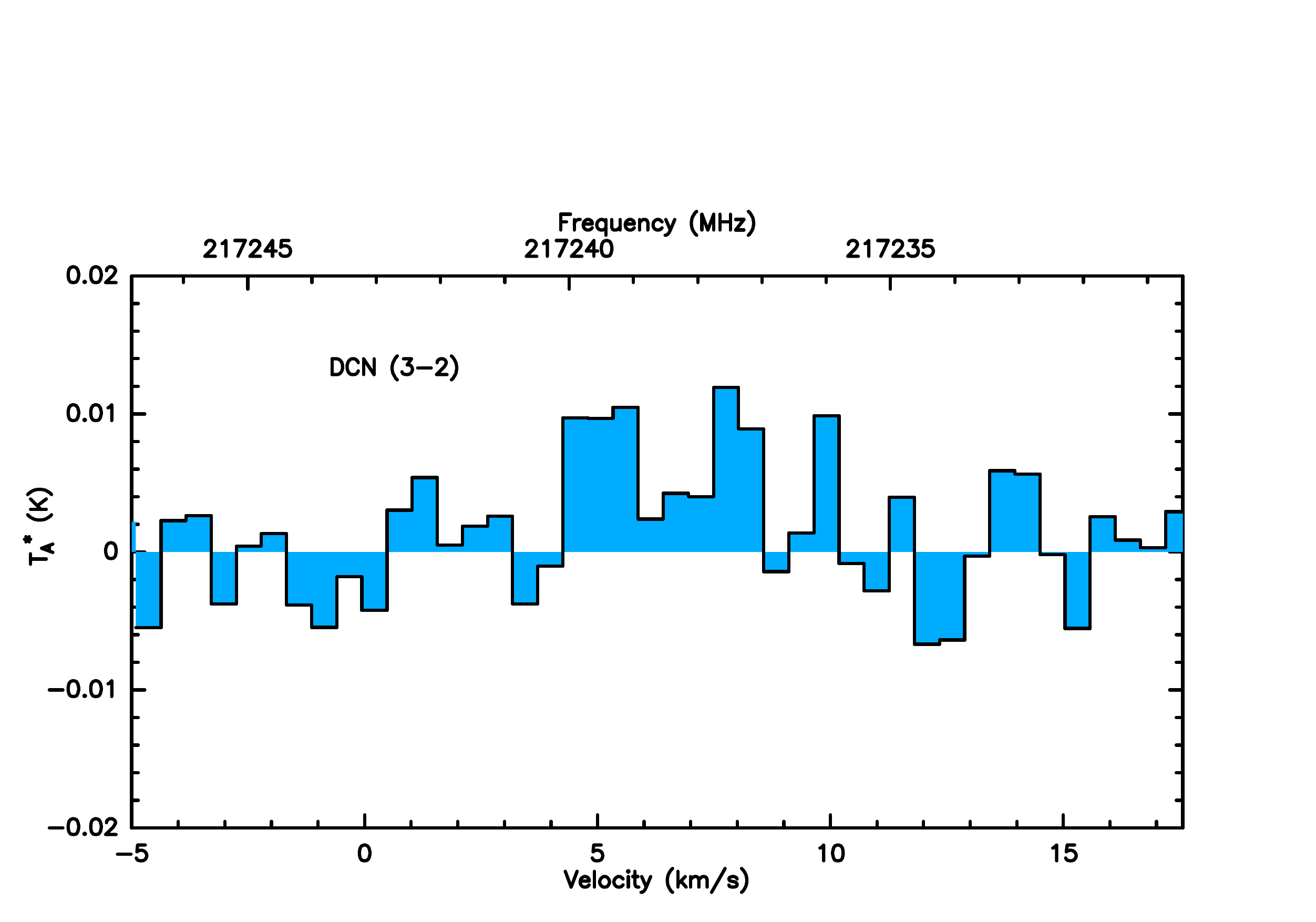}
    \includegraphics[scale=0.45, angle=270]{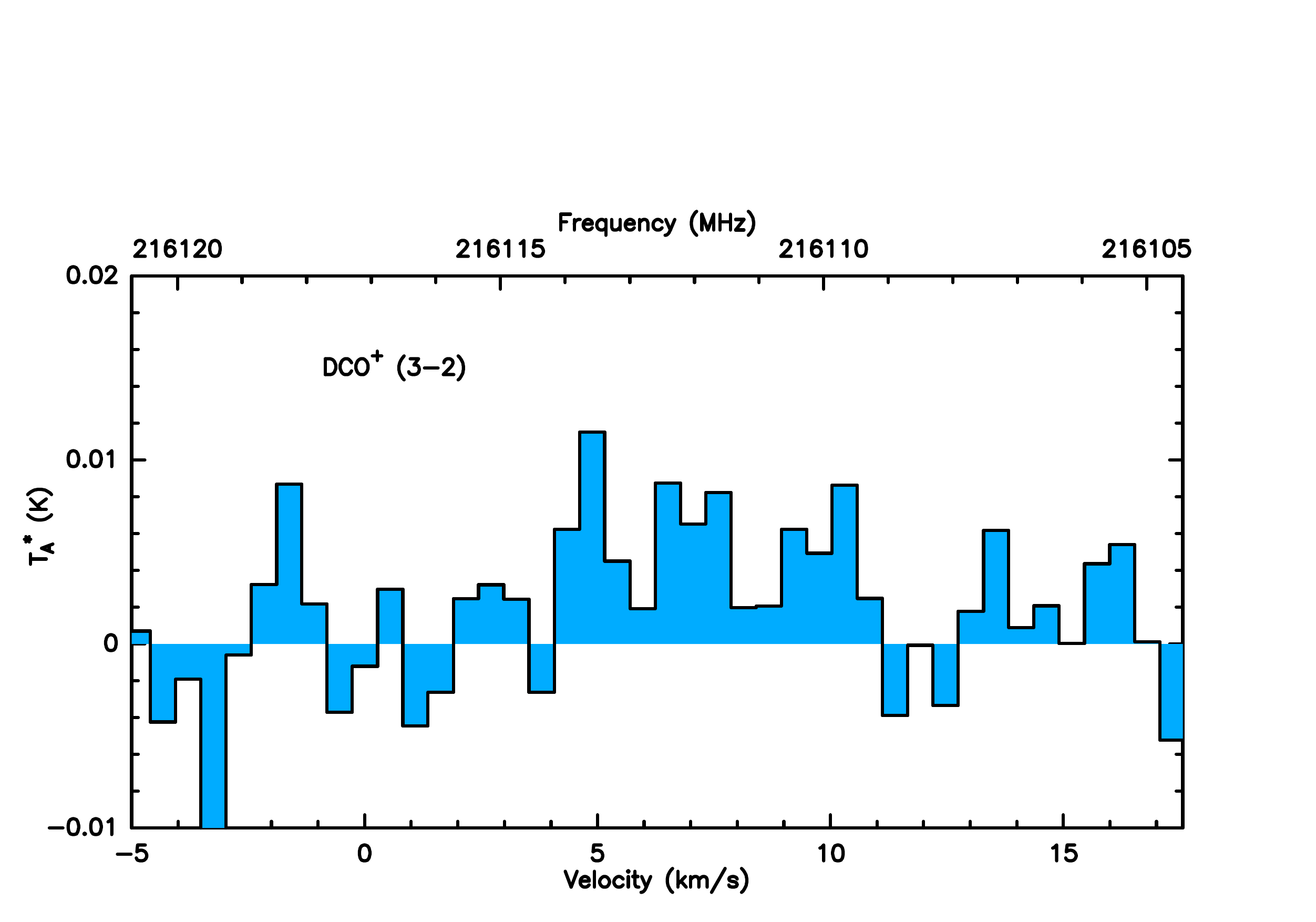}
\end{figure}

For all lines with high signal-to-noise, all four parameters were allowed to vary to find the best-fitting line profile, as determined by minimizing the reduced $\chi^{2}$ of the fit.
To fit the low signal-to-noise lines, we adopt the model parameters for $v_d$, $q$, and $p_d$ from the ${\rm ^{12}CO}$ line and only allow $T_{mb}$ to remain free.
To estimate upper limits and tentative detections, we fixed all parameters, except $T_{mb}$, to the values that best fit the ${\rm ^{12}CO}$ line, and then determined the maximum line intensity that could be present given the observed noise level.
The upper limit (at the one-sigma level) was then determined by integrating over the resulting model line profile.
The results of these line profile fits are listed in Table~\ref{tbl:MyDetections}; the upper limits listed are at the 3-$\sigma$ level.
In Figures~\ref{fig:CO}-\ref{fig:H2CO}, we display the observed and best-fit model line profiles for the transitions of CO isotopologues, ${\rm HCO^+}$, HCN, CS, and  ${\rm H_2CO}$; in all of these cases, all four model profile parameters were allowed to vary.
We also obtain tentative detections for DCN and ${\rm DCO^+}$, for which 3 of the 4 parameters in the fit were fixed.
The spectra of these tentative detections were rebinned and are presented in Figure~\ref{fig:tentativedetections}.
For comparison, a summary of previous molecular line intensity measurements for LkCa 15 is provided in Table~\ref{tbl:MolecularSpeciesALLPub}.

\subsection{Column Density Estimates}
\label{sec:CDE}

We estimate source-averaged column densities for the (pure rotational) molecular transitions detected using the methods described in \citet*{1999ApJ...517..209G} and \citet{2014ApJ...793...55K}, where we use the results for the integrated line intensities (Table~\ref{tbl:MyDetections}), along with molecular line reference data from The Cologne Database for Molecular Spectroscopy (CDMS)\footnote{\footnotesize See http://www.astro.uni-koeln.de/cdms.}.
These methods are briefly summarized here.

In the case of optically thin emission ($\tau \ll 1$), the total column density for linear molecules -- for which the degeneracy of the upper rotational level is given by $g_{u} = 2J_{u}+1$ and the energy of the lower level is $E_{l} = h B_{rot} J_{l} (J_{l}+1)$ -- can be approximated by
\begin{equation}
N_{tot} = \frac{8 \pi \nu^{3}}{c^{3}}\frac{Q_{rot}(T_{ex})}{A_{ul}g_{u}} \frac{e^{E_{l}/kT_{ex}}}{1-e^{T_{ul}/T_{ex}}} \frac{W}{\triangle J_{\nu}} f_{dilution},\\
\label{eq:columndensity}
\end{equation}
where $\nu$ is the frequency of the transition, $W$ is the integrated line intensity (in K cm s$^{-1}$), $A_{ul}$ is the Einstein coefficient for the transition, $T_{ul} = \frac{h \nu}{k}$ is the temperature equivalent of the transition energy, $f_{dilution}=\frac{\Omega_{mb}+\Omega_{s}}{\Omega_{s}}$ is the main-beam dilution correction factor (with $\Omega_{mb}$ and $\Omega_{s}$ the main-beam and source solid angles, respectively), and $\Delta J_{\nu} = J_{\nu}(T_{ex})-J_{\nu}(T_{bg})$ where $J_{\nu} = \frac{h \nu}{k} \frac{1}{exp(h \nu/kT)-1}$ and $T_{bg}$ is the temperature of the cosmic microwave background radiation.
For optically thick emission ($\tau\stackrel{>}{\sim}1$), a correction factor of $\frac{\tau}{1-e^{-\tau}}$ must be applied to Equation~\ref{eq:columndensity}.

For the simple (diatomic) case where the rotational level is not split by higher-order effects (e.g., fine, hyperfine structures) the calculation of the total column density is straightforward.
The temperature dependent total partition function then reduces to the rotational partition function, which can be approximated by
\begin{equation}
Z(T_{ex}) \approx Q_{tot}(T_{ex}) \approx Q_{rot}(T_{ex}) \approx \frac{1}{3} + \frac{kT_{ex}}{hB_{rot}},
\label{eq:partitionfunc}
\end{equation}
where $T_{ex}$ is the excitation temperature of the gas and $B_{rot}$ is the molecular rotational constant (see, e.g., \citealt{2015arXiv150101703M}).
Equation~\ref{eq:partitionfunc} only applies to the case of linear (and approximately linear) molecules, such that the molecules have only one (or only one dominant) moment of inertia, or cases in which the hyperfine structures are not resolved.
In the (more complex) case of the ${\rm H_2CO}$ molecule, we adopt the values of the partition function provided by the CDMS database and interpolate to determine the value for the partition function at arbitrary temperatures.

Using the foregoing methods, we obtain source-averaged column densities N$_{X}$ for a range of excitation temperatures ($T_{ex}$ = 4, 6, 10, 20, 30 K) for the species that are assumed to be optically thin.
Molecular data (e.g., $\nu$, $B_{rot}$, $A_{ul}$) were obtained from the CDMS database.
In Table~\ref{tbl:ColumnDensities}, we list these column densities for the foregoing excitation temperatures, along with column density estimates for ${\rm ^{13}CO}$ (2-1).
Although the ${\rm ^{13}CO}$ (2-1) emission is evidently optically thin (\S~\ref{sec:OpticalDepthApprox}), the column density estimates for ${\rm ^{13}CO}$ (2-1) were derived from the column density obtained from the ${\rm C^{18}O}$ integrated intensity, adopting isotopic abundance ratios ${\rm ^{12}C}$:${\rm ^{13}C}$ $\sim$ 68 \citep{2005ApJ...634.1126M} and ${\rm ^{16}O}$:${\rm ^{18}O}$ $\sim$ 480 \citep{2006A&A...456..675S}.

For all molecules except CN and ${\rm C_2H}$ (see \S~\ref{sec:CNC2H}), the values of column density (Table~\ref{tbl:ColumnDensities}) were obtained assuming the characterisitic emitting region is a uniform disk of angular radius $R=4''$, based on results from interferometric studies (e.g., \citealt{2007A&A...467..163P,2010ApJ...720..480O,2012A&A...537A..60C}).
Hence, we have adopted a beam dilution factor\footnote{\footnotesize In \citet{2014ApJ...793...55K}, we mistakenly quote the inverse of the beam dilution factor.} of $f_{dilution}$ $\approx$ 1.8 for all molecules in the disk, corresponding to the half power beam width of the IRAM 30 m telescope at 230 GHz (10.7'').
This simplification introduces modest systematic errors in the values of $f_{dilution}$ and, hence, column density at the extremes of the frequency range surveyed (i.e., a $\sim 6\%$ underestimate at 210 GHz and a $\sim 30\%$ overestimate at 260 GHz).

\subsection{Opacities in CO Lines}
\label{sec:OpticalDepthApprox}

We can take advantage of the detection of ${\rm C^{18}O}$ to estimate the optical depths of the ${\rm ^{13}CO}$ (2-1) and ${\rm ^{12}CO}$ (2-1) transitions.
The optical depth of ${\rm ^{13}CO}$ may be estimated by comparing the expected and measured line ratios for the CO isotopologues.
We expect
\begin{equation}
R = \frac{1-\exp{(-\tau_{^{13}CO})}}{1-\exp{(-\tau_{^{13}CO}/X)}},
\label{eq:ODApprox}
\end{equation}
where $R$ is the measured ${\rm ^{13}CO}$:${\rm C^{18}O}$ line ratio and $X$ is the ${\rm ^{13}CO}$:${\rm C^{18}O}$ abundance ratio.
Equation~\ref{eq:ODApprox} follows directly from the equation of radiative transfer, if it is assumed that the contribution from the background radiation is negligible and the excitation temperatures of ${\rm ^{13}CO}$ and ${\rm C^{18}O}$ are identical.
We adopt a value of $X=7$ for the abundance ratio, ${\rm ^{13}C}$:${\rm ^{12}C}$/${\rm ^{18}O}$:${\rm ^{16}O}$, based on the values ${\rm ^{12}C}$:${\rm ^{13}C}$ = 68 and ${\rm ^{16}O}$:${\rm ^{18}O}$ = 480 (as determined from measurements of CN and CO isotopologues for the local interstellar medium and the Solar System, respectively; \citealt{2005ApJ...634.1126M,2006A&A...456..675S}).
So, if the emisson from ${\rm ^{13}CO}$ is optically thin ($\tau_{^{13}CO}$ $\ll$ 1), we would expect to obtain $R \approx 7$.
The measured ratio for the $J=2 \rightarrow 1$ transitions for LkCa 15, $R \approx 6.0$, therefore confirms that the emission in the ${\rm ^{13}CO}$ (2-1) line is optically thin, with an implied optical depth of $\tau_{^{13}CO} \sim 0.38$.
Similar results were obtained by \citet{2003ApJ...597..986Q} from observations of LkCa 15 in the $J = 1 \rightarrow 0$ transitions of ${\rm ^{13}CO}$ and ${\rm C^{18}O}$.
Then, given the assumptions for the isotopic abundances of carbon, we expect $\tau_{^{12}CO} \approx 68 \tau_{^{13}CO}$.
Hence, we estimate an optical depth of $\sim 26$ for ${\rm ^{12}CO}$ (2-1), i.e., the emission in this transition of ${\rm ^{12}CO}$ from the LkCa 15 disk is highly optically thick.

We caution that the foregoing estimates of ${\rm ^{13}CO}$ optical depths are potentially subject to large uncertainties.
These CO opacity estimates assume that there is no selective photodissociation or chemical fractionation across the line profile.
Finally (and perhaps least significantly), the ${\rm ^{13}CO}$ line profile is slightly asymmetric (see Figure~\ref{fig:CO}), which suggests that the telescope was mispointed by a few arcseconds for that particular observation, and hence, the intensity of the $^{13}$CO line may be a slight underestimate.

\subsection{CN and ${\rm C_2H}$ Hyperfine Component Analysis: Optical Depths, Excitation Temperatures, and Column Densities}
\label{sec:CNC2H}

\begin{figure}[tbp]
  \caption{Full spectra of hyperfine components of CN observed toward LkCa 15. The red lines indicate the positions and relative intensities of the hyperfine components (See Table~\ref{tbl:CNC2HDetections}). The spectra are overlaid with the best-fit models obtained via the method described in \S~\ref{sec:CNC2H}. The residuals from the fit are below the spectra.}
 \label{fig:CN}
  \centering
    \includegraphics[scale=0.5, angle=0, width=1.0\textwidth]{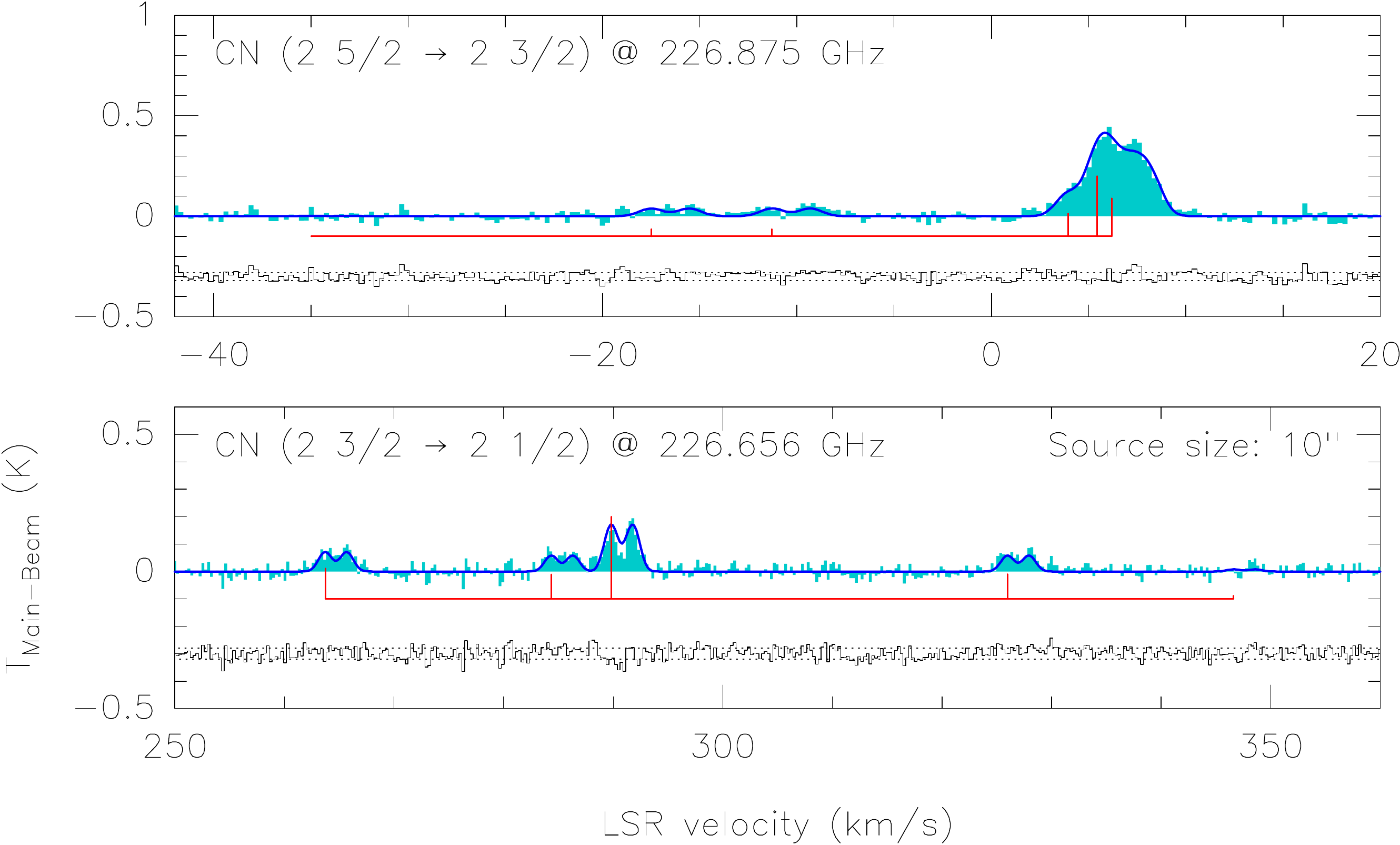}
\end{figure}

\begin{figure}[tbp]
  \caption{Full spectrum of hyperfine components of ${\rm C_2H}$ observed toward LkCa 15. The red lines indicate the positions and relative intensities of the hyperfine components (See Table~\ref{tbl:CNC2HDetections}). The spectrum is overlaid with the best-fit models obtained via the method described in \S~\ref{sec:CNC2H}. The residuals from the fit are below the spectrum.}
 \label{fig:C2H}
  \centering
 \includegraphics[scale=0.5, angle=0, width=1.0\textwidth]{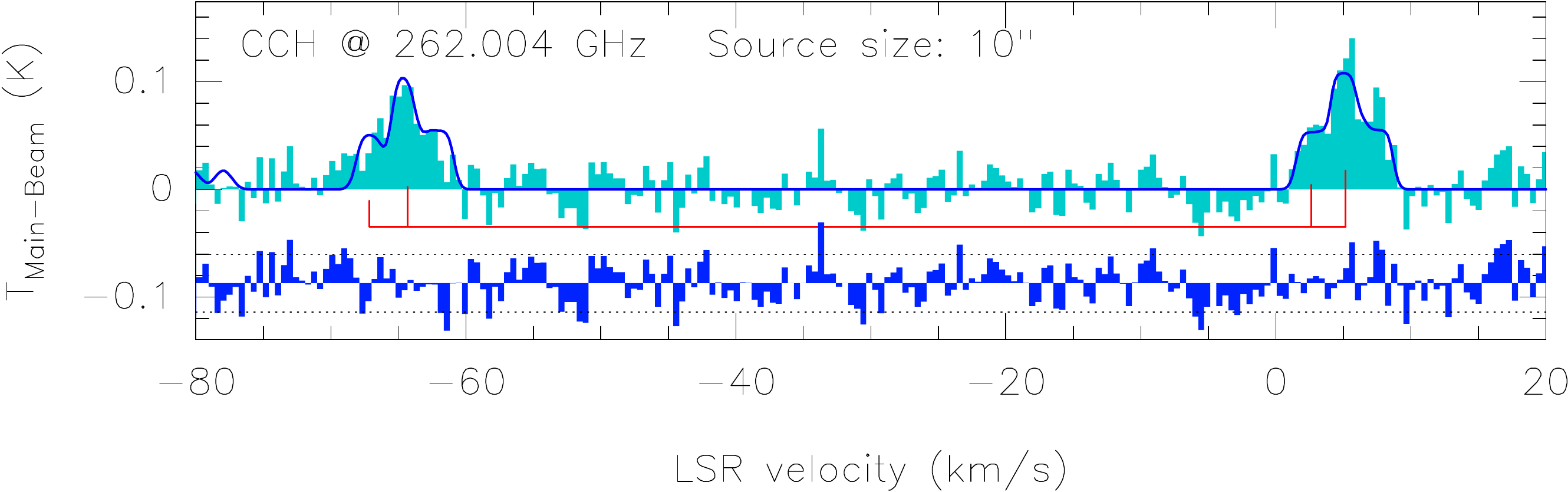}
\end{figure}

The CN and ${\rm C_2H}$ rotational emission lines display hyperfine splitting (HFS) due to interactions between electron and nuclear spins (e.g., \citealt{1982ApJ...254...94Z}, see Table~\ref{tbl:CNC2HDetections}).
We performed measurement and analysis of the LkCa 15 CN $N =2 \rightarrow 1$ and ${\rm C_2H}$ $N =3 \rightarrow 2$ spectra --- which are displayed in Figure~\ref{fig:CN} and Figure~\ref{fig:C2H}, respectively --- using a new adaptation of the classical HFS fitting method not yet implemented in the CLASS software.
This adaptation uses two Gaussians of the same width to fit the double-peaked emission features in our spectrum that result from the inclined Keplerian disk.
The two-Gaussian HFS code makes the simplifying assumption that all the hyperfine transitions within a given rotational transition share the same excitation temperature.
Fitting all the hyperfine components simultaneously results in the determination of the opacity, $\tau$, and excitation temperature, $T_{ex}$, of the rotational transition \citep[see][]{2014ApJ...793...55K}.
Figure~\ref{fig:CN}, Figure~\ref{fig:C2H}, and Table~\ref{tbl:hfs} summarize the results of this analysis for the IRAM 30m measurements of the CN and ${\rm C_2H}$ emission from LkCa 15.

The determination of the total partition functions $Q_{tot}$ for the CN and ${\rm C_2H}$ radicals must account for the nuclear-spin statistics, as discussed in \citet{2014ApJ...793...55K}.
The column density is then computed as
\begin{equation}
N_{tot} = \frac{8 \pi \nu^{3}}{c^{3}} \frac{e^{E_{l}/kT_{ex}}}{1-e^{-T_{ul}/T_{ex}}} \int{\tau_{ul} d\nu}\frac{Q_{tot}(T_{ex})}{A_{ul}g_{u}},
\label{eq:HFSCDE}
\end{equation}
where all parameters refer to the particular hyperfine transition of interest.
We assume the characteristic emitting region is a uniform disk of angular radius $R=5''$, based on results from interferometric studies (e.g., \citealt{2012A&A...537A..60C,2010ApJ...720..480O}).
We adopt a beam dilution correction factor $f_{dilution} \approx 1.1$, but the results we obtain are fairly insensitive to the assumed source solid angle.
Following this procedure, we obtain values of $T_{ex}\sim 3.9$ K and $\tau \sim 2.2$ for CN, and $T_{ex} \sim 2.9$ and  $\tau \sim 18$ for ${\rm C_2H}$, leading to estimates of N$_{tot}$(CN) $\sim 0.5 \times 10^{14}$ cm$^{-2}$ and N$_{tot}$(${\rm C_2H}$) $\sim 2 \times 10^{16}$ cm$^{-2}$ for the LkCa 15 disk (Table~\ref{tbl:hfs}).
The results for $\tau$ indicate the emission in both lines is optically thick, where the reader is reminded that the resulting values of $\tau$ represent the total opacity over the HFS.
Hence, the derived values of $T_{ex}$ imply either that the gas is subthermally excited or that the CN and ${\rm C_2H}$ emission originates from disk gas at very low temperature.
We caution, however, that the low $T_{ex}$ for CN and ${\rm C_2H}$ rest on the assumed beam filling factor.
In addition, a less extended disk in these molecules would imply a higher $T_{ex}$.

\section{Discussion}

In Table~\ref{tbl:ColumnDensities}, we list fractional molecular abundances relative to ${\rm ^{13}CO}$, N(X)/N(${\rm ^{13}CO}$), as obtained from the calculated source-averaged molecular column densities for LkCa 15 (\S~\ref{sec:CDE}).
We present values of N(X)/N(${\rm ^{13}CO}$) that we obtained by adopting $T_{ex}$ = 4, 6, 10, 20, 30 K.
We caution that these column density estimates may have large systematic uncertainties, given the uncertainties in the assumed source parameters and, in particular, the assumed source emitting region solid angles in the various molecules \citep{2007A&A...467..163P,2010ApJ...720..480O,2003PASJ...55...11A,2012A&A...537A..60C}.
Nevertheless, our results yield column densities that are overall consistent with previous studies of the LkCa 15 disk.
This consistency is apparent from Table~\ref{tbl:ColumnDensityComparison}, in which we present a comparison of our column densities and column densities determined by previous studies.
In most cases, our column density estimates agree with previous estimates to within a factor $\sim$3.
Apart from CN and ${\rm C_2H}$ (see below), the largest discrepancies appear in the comparison with results for ${\rm ^{13}CO}$, ${\rm C^{18}O}$, and ${\rm HCO^+}$ published by \citet{2003ApJ...597..986Q}; the large differences between column density results in these cases can be attributed to the constrasting beam sizes of OVRO ($\sim 3"$) and IRAM ($\sim 12"$).

It should be noted that for LkCa 15, we derive column densities for CN and especially for ${\rm C_2H}$ that are relatively high; in the latter case, the inferred source-averaged column density is similar to that of ${\rm ^{13}CO}$.
Specifically, we derive N(CN)/N(${\rm ^{13}CO}$) and N(${\rm C_2H}$)/N(${\rm ^{13}CO}$) values of approximately 0.01 and 3.4, respectively, adopting $T_{ex}$ = 4 K.
The former result is somewhat smaller than that quoted for LkCa 15 in \citet{2014ApJ...793...55K}, where the difference reflects our refined treatment of CN optical depth and excitation temperature (\S~\ref{sec:CNC2H}), as well as different assumptions regarding beam filling factors.
In deriving our column densities, we assumed that all the molecular emission is distributed within a smooth disk, but clearly this may not be the case.
If the CN or ${\rm C_2H}$ emission were instead restricted to specific regions of the disk, we could be severely underestimating the column densities of these molecular species.
Thus, it is imperative to obtain high signal-to-noise interferometric images to study the surface brightness distribution for these highly abundant molecules.

It is possible that high-energy (FUV and/or X-ray) radiation from the central star may be enhancing the abundances of CN and ${\rm C_2H}$ in T Tauri disks (\citealt{1997Sci...277...67K,2008A&A...492..469K,2014ApJ...793...55K,1997A&A...317L..55D,2004A&A...425..955T,2010ApJ...714.1511H}).
Suggestions about the influence of X-rays on disk chemistry were made by \citet{2008A&A...492..469K}, \citet{2004A&A...425..955T}, \citet{2011A&A...536A..80S} to explain enhanced HCN and ${\rm HCO^+}$ line fluxes, and by \citet{2014ApJ...793...55K} to explain the enhanced abundances that were found for CN and ${\rm C_2H}$ for TW Hya and V4046 Sgr.

We also find very low values of excitation temperature for CN and ${\rm C_2H}$ ($T_{ex} \sim$ 3-5 K) in the LkCa 15 disk.
These determinations of low $T_{ex}$ may indicate that the emission from CN and ${\rm C_2H}$ arises from regions that are deep within the disk.
Hence, as noted by \citet{2014ApJ...793...55K}, it may be that the abundances of CN and ${\rm C_2H}$ are being enhanced by X-rays as opposed to extreme FUV photons, since X-rays can penetrate deeper into disks (e.g., \citealt*{2013ApJ...765....3S,2013ApJ...772....5C}).
Indeed, \citet{2013ApJ...765....3S} found that LkCa 15 is an X-ray luminous source, with a spectral energy distribution that contains contributions from both soft and hard components.
This combination may increase the abundances of species (like ${\rm C_2H}$) that are the dissociation products of complex organic molecules.
Alternatively, the CN and ${\rm C_2H}$ molecules may reside in tenuous regions of the disk where collision probabilities are low, such that the molecules are subthermally excited.

Our estimates of N$_{tot}$(CN) and N$_{tot}$(${\rm C_2H}$) for the LkCa 15 disk (Table~\ref{tbl:hfs}) are similar to those found for the TW Hya and V4046 Sgr disks (\citealt{2014ApJ...793...55K}, their Table 5).
Hence, it appears that large relative abundances of ${\rm C_2H}$ (as well as CN; see \citealt{2012A&A...537A..60C}) may be a common feature of evolved protoplanetary disks.
The excitation temperatures that we find for CN and ${\rm C_2H}$ ($T_{ex} \sim$ 3-5 K) in the LkCa 15 disk are similar to, though even lower than, those determined by \citet{2014ApJ...793...55K} for TW Hya and V4046 Sgr; futhermore, we have inferred relatively large values for the optical depths of CN and ${\rm C_2H}$ for all three of these disks.
Subsequently, Submillimeter Array imaging of TW Hya revealed that ${\rm C_2H}$ emission emenates from a ring-like structure, with an inner radius ($\sim45$ AU) that appears to lie between the CO ``snow line'' (as traced by the $\sim30$ AU-diameter hole in ${N_2H^+}$) and the outer boundary of large-grain emission at $\sim60$ AU \citep{Kastner2015}.
Given the similarity of the TW Hya and LkCa 15 disks in terms of the excitation and optical depth of ${\rm C_2H}$, it is likely that the ${\rm C_2H}$ (and possibly CN) emission from LkCa 15 also originates from a ring-like structure.

We present fractional abundances with respect to ${\rm ^{13}CO}$, N(X)/N(${\rm ^{13}CO}$) in Table~\ref{tbl:fracabun}, where we cite the results of \citealt{2004A&A...425..955T} as a cross-check of our results.
This table illustrates that CN and ${\rm C_2H}$ are enhanced in evolved T Tauri star disks relative to other astronomical objects, especially with respect to YSOs and even the expanding, C-rich envelope of the classical carbon star IRC+10216, in which the abundances of carbon-bearing molecules are strongly enhanced.
Thus, our line survey of LkCa 15 reinforces previous assertions (e.g., \citealt{2014ApJ...793...55K} and references therein) that there must be some mechanism that is enhancing the abundances of CN and ${\rm C_2H}$ in protoplanetary disks.

\section{Summary}

We have used the Institute de Radioastronomie Millim\'{e}trique (IRAM) 30 meter telescope and the EMIR fast fourier transform spectrometers to conduct a comprehensive mm-wave emission line survey of the circumstellar disk orbiting the nearby, pre-main sequence (T Tauri) star LkCa 15 over the 210-267 GHz (1.4 - 1.1 mm) range.
We find that lines of ${\rm ^{12}CO}$, ${\rm HCO^+}$, HCN,  ${\rm ^{13}CO}$, CN, ${\rm C_2H}$, CS, ${\rm H_2CO}$, and ${\rm C^{18}O}$ constitute the strongest molecular emission from the LkCa 15 disk in the spectral region surveyed.
The $J=5 \rightarrow 4$ transition of CS and the ($J=3_{13}\rightarrow2_{12}$) transition of ${\rm H_2CO}$ are detected for the first time in the LkCa 15 disk.
We use simultaneous measurements of CO and its isotopologues (i.e., ${\rm ^{13}CO}$ and ${\rm C^{18}O}$) to estimate the optical depths of the transitions of CO and ${\rm ^{13}CO}$ observed; we find that ${\rm ^{13}CO}$ emission is optically thin, and emission from ${\rm ^{12}CO}$ is highly optically thick.
The survey also includes first measurements of the full suite of hyperfine transitions of CN $N = 2 \rightarrow 1$ and ${\rm C_2H}$ $N = 3 \rightarrow 2$.
Modeling of these CN and ${\rm C_2H}$ hyperfine complexes indicates that the emission from both species is optically thick and suggests either that the emission from these molecules originates from either very cold ($<$10 K) regions of the disk, or that the molecules reside in less dense upper atmosphere regions of the disk and are subthermally excited.
We also find that the column densities for CN and ${\rm C_2H}$ are comparable to those of isotopologues of CO, implying that these molecules are very abundant in the disk of LkCa 15.

As with our molecular line surveys of TW Hya and V4046 Sgr \citep{2014ApJ...793...55K}, the results of our molecular line survey of LkCa 15 stress the need for additional molecular line surveys of protoplanetary disks that cover a range of pre-MS evolutionary states; interferometric imaging of protoplanetary disks in transitions of CN and ${\rm C_2H}$, as well as other potential molecular tracers of UV and/or X-ray radiation; and detailed modeling of T Tauri star-disk systems that includes the effects of high-energy radiation on disk chemistry and is constrained by observations of pre-MS UV and X-ray fields.
In general, we note that interferometric spectral surveys will become very important in the upcoming years.
Unbiased line surveys carried out with interferometers would allow for the simultaneous determination of the chemical inventory of disks and the emitting regions sizes and morphologies of these molecules.

\acknowledgments{\it This work is based on observations carried out with the IRAM 30m Telescope. IRAM is supported by INSU/CNRS (France), MPG (Germany) and IGN (Spain). This research is supported by National Science Foundation grant AST-1108950 to RIT. The authors thank Charlie Qi for helpful discussions.}



\bibliographystyle{apj}



\thispagestyle{empty}

\begin{deluxetable}{c c c c c c}

\centering
\tablecolumns{6}
\tablewidth{340pt}
\tabletypesize{\footnotesize}

\tablecaption{\label{tbl:MyDetections}\sc LkCa 15: Molecular Species Detected in the IRAM 30 m Line Survey}

\tablehead{

	\multicolumn{1}{c}{Species} &
	\multicolumn{1}{c}{Transition} &
	\multicolumn{1}{c}{$\nu$} &
	\multicolumn{1}{c}{T$_{mb}$} &
	\multicolumn{1}{c}{$v_{d}$ $^{a}$} &
	\multicolumn{1}{c}{$I$} \\

	\multicolumn{1}{c}{} &
	\multicolumn{1}{c}{} &
	\multicolumn{1}{c}{(GHz)} &
	\multicolumn{1}{c}{(K)} &
	\multicolumn{1}{c}{(km s$^{-1}$)} &
	\multicolumn{1}{c}{(K km s$^{-1}$) $^{f}$} \\
}

\startdata

$^{12}$CO & $J=2\rightarrow 1$ &  230.5380000 & 0.542 & 1.128 & 1.441 \\
$^{13}$CO &  $J=2\rightarrow 1$ & 220.3986765 & 0.201 & 1.184 & 0.550 \\
C$^{18}$O &  $J=2\rightarrow 1$ & 219.5603568 & 0.038 & 1.124 & 0.092 \\
CN & $N=2\rightarrow 1$ & 226.6321901 & 0.036 & 1.413 & 0.108 \\
       &  & 226.6595584 & 0.079 & 1.084 & 0.226 \\ 
       &  & 226.6636928 & 0.040 & 1.228 & 0.118 \\
       &  & 226.6793114 & 0.042 & 1.251 & 0.112 \\
       &  & 226.8741908 & 0.183 & 1.431 & 0.681$^{b}$ \\
       &  & 226.8747813 & \ldots & \ldots & \ldots \\
       &  & 226.8758960 & \ldots & \ldots & \ldots \\
       &  & 226.8874202 & 0.020 & 1.762 & 0.081 \\
       &  & 226.8921280 & 0.015 & 1.188 & 0.062 \\
CS &  $J=5\rightarrow 4$ & 244.9356435 & 0.064 & 1.245 & 0.182 \\ 
C$_{2}$H & $N=3\rightarrow 2$ & 262.0042266 & 0.049 & 1.128 & 0.269$^{c}$ \\
    & & 262.0064034 & \ldots & \ldots & \ldots \\
    &  & 262.0648433 & 0.033 & 1.128 & 0.152$^{d}$ \\
    &  & 262.0673312 & \ldots & \ldots & \ldots \\
HCN & $J=3\rightarrow 2$ & 265.8861800 & 0.238 & 1.411 & 0.732 \\           
HCO$^{+}$ & $J=3\rightarrow 2$ & 267.5575260 & 0.334 & 1.298 & 1.024 \\
DCN & $J=3\rightarrow 2$ & 217.2386307 & 0.019 & 1.128 & 0.152$^{e}$ \\
DCO$^{+}$ & $J=3\rightarrow 2$ & 216.1125766 & 0.015 & 1.128 & 0.118$^{e}$ \\
H$_2$CO & $J=3_{13}\rightarrow 1_{12}$ & 211.2114680 & 0.056 & 1.300 & 0.147 \\
        & $J=3_{03}\rightarrow 2_{02}$ & 218.2221920 & 0.032 & 1.134 & 0.085 \\
        & $J=3_{22}\rightarrow 2_{21}$ & 218.4756320 & $<$0.008 & 1.128 & $<$0.060 \\
        & $J=3_{12}\rightarrow 2_{11}$ & 225.6977750 & 0.058 & 1.093 & 0.136 \\

\enddata

\begin{flushleft}
a) As defined by \citet{2008A&A...492..469K}.\\
b) Sum of integrated intensities of hyperfine structure lines in range 226.87419-226.87590 GHz.\\
c) Sum of integrated intensities of hyperfine structure lines in range 262.00426-262.00649 GHz.\\
d) Sum of integrated intensities of hyperfine structure lines in range 262.06498-262.06747 GHz.\\
e) Tentative detection.\\
f) These line intensities can be converted to Jy km s$^{-1}$ using $3.906 \frac{F_{eff}}{A_{eff}}$ Jy/K for $S/T_{A}^{*}$, the point source sensitivity, with $F_{eff} = 0.92$ and $A_{eff} = 0.46$ (aperture efficiency), where these represent mean values for the IRAM 30 m telescope in the 1.3 mm window.\\
\end{flushleft}
\end{deluxetable}



\newgeometry{bottom=1cm,left=1cm,right=1cm}

\begin{landscape}
\begin{deluxetable}{c l c c c c c}

\centering
\tablecolumns{7}
\tablewidth{600pt}
\tabletypesize{\footnotesize}

\tablecaption{\label{tbl:CNC2HDetections}\sc Transitions of CN and C$_{2}$H}

\tablehead{

	\multicolumn{1}{c}{Species} &
	\multicolumn{1}{c}{Transition} &
	\multicolumn{1}{c}{$\nu$ $^{a}$} &
	\multicolumn{1}{c}{$A_{ul}$ $^{a}$} &
	\multicolumn{1}{c}{$g_{u}$ $^{a}$} &
	\multicolumn{1}{c}{Relative Intensities $^{a}$} &
	\multicolumn{1}{c}{$I$} \\

	\multicolumn{1}{c}{(Rotational Transition)} &
	\multicolumn{1}{c}{} & 
	\multicolumn{1}{c}{(GHz)} &
	\multicolumn{1}{c}{(s$^{-1}$} &
	\multicolumn{1}{c}{} &
	\multicolumn{1}{c}{} &
	\multicolumn{1}{c}{(K km s$^{-1}$)} \\
}

\startdata

CN 
 & $J=3/2 \rightarrow 1/2, F=3/2 \rightarrow 3/2$ & 226.6321901 & $4.26\times10^{-5}$ & 4 & 0.0533 & 0.106 \\
 ($N=2\rightarrow 1$) & $J=3/2 \rightarrow 1/2, F=5/2 \rightarrow 3/2$ & 226.6595584 & $9.47\times10^{-5}$ & 6 & 0.1776 & 0.221 \\
   & $J=3/2 \rightarrow 1/2, F=1/2 \rightarrow 1/2$ & 226.6636928 & $8.47\times10^{-5}$ & 2 & 0.0529 & 0.116 \\
   & $J=3/2 \rightarrow 1/2, F=3/2 \rightarrow 1/2$ & 226.6793114 & $5.27\times10^{-5}$ & 4 & 0.0659 & 0.110 \\
   & $J=5/2 \rightarrow 3/2, F=5/2 \rightarrow 3/2$ & 226.8741908 & $9.62\times10^{-5}$ & 6 & 0.1805 & 0.667$^{b}$ \\
   & $J=5/2 \rightarrow 3/2, F= 7/2\rightarrow 5/2$ & 226.8747813 & $1.14\times10^{-4}$ & 8 & 0.2860 & \ldots \\
   & $J= 5/2\rightarrow 3/2, F= 3/2\rightarrow 1/2$ & 226.8758960 & $8.59\times10^{-5}$ & 4 & 0.1074 & \ldots \\
   & $J= 5/2\rightarrow 3/2, F= 3/2\rightarrow 3/2$ & 226.8874202 & $2.73\times10^{-5}$ & 4 & 0.0342 & 0.080 \\
   & $J= 5/2\rightarrow 3/2, F=5/2 \rightarrow 5/2$ & 226.8921280 & $1.81\times10^{-5}$ & 6 & 0.0340 & 0.060 \\
\hline
C$_{2}$H 
& $J=7/2 \rightarrow 5/2, F= 4\rightarrow3$ & 262.0042600 & $5.32\times10^{-5}$ & 9 & 0.3213 & 0.284$^{c}$ \\
($N=3\rightarrow2$) & $J=7/2 \rightarrow 5/2, F= 3\rightarrow2$ & 262.0064820 & $5.12\times10^{-5}$ & 7 & 0.2404 & \ldots \\
& $J=5/2 \rightarrow 3/2, F= 3\rightarrow2$ & 262.0649860 & $4.89\times10^{-5}$ & 7 & 0.2296 & 0.160$^{d}$ \\
& $J=5/2 \rightarrow 3/2, F= 2\rightarrow1$ & 262.0674690 & $4.47\times10^{-5}$ & 5 & 0.1501 & \ldots \\

\enddata

\begin{flushleft}
a) Values of frequencies, $A_{ul}$, $g_{u}$, and theoretical relative intensities of hyperfine transitions of CN and C$_{2}$H obtained from the CDMS database. CN line frequencies were measured in the laboratory by \citet{skatrud1983} and the fitted values from the CDMS database are from \citet{2000A&A...357L..65M}.\\
b) Sum of integrated intensities of hyperfine structure lines in range 226.87419-226.87590 GHz.\\
c) Sum of integrated intensities of hyperfine structure lines in range 262.00426-262.00649 GHz.\\
d) Sum of integrated intensities of hyperfine structure lines in range 262.06498-262.06747 GHz.\\
\end{flushleft}
\end{deluxetable}
\end{landscape}


\newgeometry{left=1cm,right=1cm}
\begin{landscape}
\begin{deluxetable}{c c c c c c}

\tablecolumns{6}
\tablewidth{520pt}
\tabletypesize{\footnotesize}

\tablecaption{\label{tbl:MolecularSpeciesALLPub}\sc LkCa 15: Molecular Transitions Observed to Date}

\tablehead{

	\colhead{Species} &
	\colhead{Transition} &
	\colhead{$\nu$} &
	\colhead{$I^{d}$} &
	\colhead{$I$} &
	\colhead{References$^{a}$} \\

	\colhead{} &
	\colhead{} &
	\colhead{(GHz)} &
	\colhead{(K km s$^{-1}$)$^{b}$} &
	\colhead{(Jy km s$^{-1}$)$^{e}$} &
	\colhead{} \\
}

\startdata

$^{12}$CO & $J=2\rightarrow 1$ &  230.5380000 & 1.42, 1.82 & 13.94, \ldots$^{g}$, 12.5, \ldots$^{g}$, \ldots$^{g}$ & 1, 2, 3, 9, 10, 11, 12 \\
& $J=3\rightarrow 2$ &  345.7959899 & 1.17, \ldots$^{n}$ & & 2, 13 \\
& $J=6\rightarrow 5$ & 691.4730763 & 1.9 & & 13 \\
$^{13}$CO & $J=1\rightarrow 0$ & 110.2013541 & & \ldots$^{g}$, 6.39 & 9, 10 \\
&  $J=2\rightarrow 1$ & 220.3986765 & 0.53 & \ldots$^{g}$ & 1, 9 \\
& $J=3\rightarrow 2$ & 330.5879601 & 0.39, \ldots$^{n}$ & & 2, 13 \\
C$^{18}$O & $J=1\rightarrow 0$ & 109.7821734 & & 1.90 & 10 \\
          & $J=2\rightarrow 1$ & 219.5603568 & 0.09, $<$0.20 & & 1, 2 \\
          & $J=3\rightarrow 2$ & 329.3305453 & $<$0.14 & & 2 \\
CN & $N=2\rightarrow 1$ & 226.6321901 & 0.11 & & 1 \\
       &  & 226.6595584 & 0.22 & & 1 \\ 
       &  & 226.6636928 & 0.12 & & 1 \\
       &  & 226.6793114 & 0.11 & 6.80 & 1, 3 \\
       &  & 226.8741908 & 0.67$^{i}$ & & 1 \\
       &  & 226.8747813 & \ldots$^{i}$ & 1.16, \ldots$^{g}$ & 1, 3, 6\\
       &  & 226.8758960 & \ldots$^{i}$ & & 1 \\
       &  & 226.8874202 & 0.08 & & 1 \\
       &  & 226.8921280 & 0.06 & & 1 \\
       & $J=3\frac{7}{2}\rightarrow 2\frac{5}{2}$ & 340.2485764 & 0.67 & & 2 \\
CS & $J=3\rightarrow 2$ & 146.9690287 & \ldots$^{g}$ & & 15 \\
        &  $J=5\rightarrow 4$ & 244.9356435 & 0.18 & & 1 \\ 
        & $J=7\rightarrow 6$ & 342.8830000 & $<$0.08$^{h}$ & & 2 \\
CCS & $J=7,7\rightarrow 6,8$ & 90.6863810 & $<$0.010$^{h}$ & & 7 \\
    & $J=7,8\rightarrow 6,7$ & 93.8701070 & $<$0.008$^{h}$ & & 7 \\
    & $J=11,12\rightarrow 10,11$ & 144.2448364 & 0.061 & & 7 \\
C$_{2}$H & $J=1\rightarrow 0$ & 87.3286240 & & \ldots$^{g}$ & 4 \\
    & $J=2\rightarrow 1$ & 174.6632220 & & \ldots$^{g}$ & 4 \\
    & $J=2\rightarrow 1$ & 174.6676850 & & \ldots$^{g}$ & 4 \\
    & $N=3\rightarrow 2$ & 262.0042266 & 0.28$^{j}$ & & 1 \\
    & & 262.0064034 & \ldots$^{j}$ & & 1 \\
    &  & 262.0648433 & 0.16$^{k}$ & & 1 \\
    &  & 262.0673312 & \ldots$^{k}$ & & 1  \\
HCN & $J=1\rightarrow0$ & 88.6318470 & & \ldots$^{g}$ & 6 \\ 
          & $J=3\rightarrow 2$ & 265.8861800 & 0.78 & 5.52 & 1, 3 \\ 
          & $J=4\rightarrow 3$ & 354.5054759 & 0.25 & & 2 \\          
HCO$^{+}$ & $J=1\rightarrow 0$ & 89.1885230 & & \ldots$^{g}$, 3.30, \ldots$^{g}$, \ldots$^{g}$ & 9, 10, 11, 12 \\
& $J=3\rightarrow 2$ & 267.5575260 & 1.10 & 5.19 & 1, 3 \\
                    & $J=4\rightarrow 3$ & 356.7341340 & 0.26 & & 2 \\
H$^{13}$CO$^{+}$ & $J=1\rightarrow 0$ & 86.7542884 & & $<$0.88 & 10 \\
& $J=4\rightarrow 3$ & 346.9983360 & $<$0.13 & & 2 \\
DCN & $J=3\rightarrow 2$ & 217.2386307 & $<$0.15 & 0.41 & 1, 3 \\
DCO$^{+}$ & $J=3\rightarrow 2$ & 216.1125766 & $<$0.11 & 0.51 & 1, 3 \\
        & $J=5\rightarrow 4$ & 360.1697771 & $<$0.10 & & 2 \\
N$_{2}$H$^{+}$ & $J=1_{1}\rightarrow 0_{1}$ & 93.1718800 & & 0.15$^{l}$, 3.83$^{l}$ & 8, 10 \\
             & $J=1_{2}\rightarrow 0_{1}$ & 93.1737000 & & \ldots$^{l}$ & 8 \\
             & $J=1_{0}\rightarrow 0_{1}$ & 93.1761300 & & \ldots$^{l}$ & 8 \\
             & $J=3\rightarrow 2$  & 279.5117010 & & 0.71 & 3 \\
					   & $J=4\rightarrow 3$ & 372.6725090 & $<$0.10 & & 2 \\
H$_{2}$D$^{+}$ & $J=1_{10}\rightarrow 1_{11}$ & 372.4213850 & $<$0.10 & & 2 \\
H$_{2}$CO & $J=2_{12}\rightarrow 1_{11}$ & 140.8395020 & 0.17 & & 2 \\
				& $J=3_{13}\rightarrow 2_{12}$ & 211.2114680 & 0.14 & & 1 \\
        & $J=3_{03}\rightarrow 2_{02}$ & 218.2221920 & 0.08, 0.14 & 0.66 & 1, 2, 3 \\
        & $J=3_{22}\rightarrow 2_{21}$ & 218.4756320 & $<$0.06, $<$0.10 & & 1, 2 \\
        & $J=3_{12}\rightarrow 2_{11}$ & 225.6977750 & 0.13, 0.10  & & 1, 2 \\
        & $J=4_{31}\rightarrow 3_{30}$ & 291.3804880 & & 1.12 & 3 \\
        & $J=5_{15}\rightarrow 4_{14}$ & 351.7686450 & 0.29 & & 2 \\
HC$_{3}$N & $J=10\rightarrow 9$ & 90.9790230 & $<$0.010$^{h}$ & & 7 \\
        & $J=12\rightarrow 11$ & 109.1736340 & $<$0.015$^{h}$ & & 7 \\
        & $J=16\rightarrow 15$ & 145.5609460 & 0.017$^{h}$ & & 7 \\
CH$_{3}$OH & $J=2_{K}\rightarrow 1_{K}$ & 96.741371 & $<$0.05 & & 2 \\
					 & $J=4_{2}\rightarrow 3_{1}$ E$^{+}$ & 218.440063 & $<$0.10 & & 2 \\
					 & $J=5_{K}\rightarrow 4_{K}$ & 241.791352 & $<$0.10 & & 2 \\
c-C$_{3}$H$_{2}$ & $J=2_{1,2}\rightarrow 1_{0,1}$ & 85.3388930 & \ldots$^{f}$ & & 5 \\
                                  & $J=6_{1,6}\rightarrow 5_{0,5}$ & 217.8221480 & \ldots$^{f}$ & & 5 \\
                                  & $J=6_{0,6}\rightarrow 5_{1,5}$ & 217.8221480 & \ldots$^{f}$ & & 5 \\
SO & $J=2_{23}\rightarrow1_{12}$ & 99.2998700 & \ldots$^{f}$ & & 15 \\
       & $J=3_{4}\rightarrow2_{3}$ & 138.1786000 & \ldots$^{f}$ & & 5 \\  
       & $J=5_{6}\rightarrow4_{5}$ & 219.9494420 & \ldots$^{f}$ & & 5 \\   
H$_{2}$S & $J=1_{10}\rightarrow1_{01}$ & 168.7627624 & \ldots$^{f}$ & & 15 \\

\enddata

\tiny
\begin{flushleft}
a) References: 1. this work; 2. \citet{2004A&A...425..955T}; 3. \citet{2010ApJ...720..480O}; 4. \citet{2010ApJ...714.1511H}; 5. \citet{2010A&A...524A..19F}; 6. \citet{2012A&A...537A..60C}; 7. \citet{2012ApJ...756...58C}; 8. \citet{2007A&A...464..615D}; 9. \citet{2007A&A...467..163P}; 10. \citet{2003ApJ...597..986Q}; 11. \citet{2000A&A...355..165D}; 12. \citet{2000ApJ...545.1034S}; 13. \citet{2001ApJ...561.1074T}; 14. \citet{2001A&A...377..566V}; 15. \citet{2011A&A...535A.104D}\\
b) Integrated line intensities are quoted in main beam temperature scale using $T_{mb} = \frac{F_{eff}}{B_{eff}}T_{A}^{*}$. See http://www.iram.es/IRAMES/mainWiki/Iram30mEfficiencies.  \\
c) Integrated line intensities for hyperfine complexes centered at the listed frequencies (see Table~\ref{tbl:CNC2HDetections}). \\
d) These values can be converted to Jy km s$^{-1}$ using $3.906 \frac{F_{eff}}{A_{eff}}$ Jy/K for $S/T_{A}^{*}$, the point source sensitivity, with $F_{eff} = 0.92$ and $A_{eff} = 0.46$. See http://www.iram.es/IRAMES/mainWiki/Iram30mEfficiencies. \\
e) Integrated line intensities quoted in Jy km s$^{-1}$ are for measurements by interferometers. \\
f) These transitions were searched for, but not detected. \\
g) These transitions were detected, but line intensities were not tabulated in the given references. \\
h) These values were converted from Jy km s$^{-1}$ using $3.906 \frac{F_{eff}}{A_{eff}}$ Jy/K for $S/T_{A}^{*}$, the point source sensitivity, with $F_{eff} = 0.95$ and $A_{eff} = 0.63$. See http://www.iram.es/IRAMES/mainWiki/Iram30mEfficiencies. \\
i) Sum of integrated intensities of hyperfine structure lines in range 226.87419-226.87590 GHz. \\
j) Sum of integrated intensities of hyperfine structure lines in range 262.00426-262.00649 GHz. \\
k) Sum of integrated intensities of hyperfine structure lines in range 262.06498-262.06747 GHz. \\
l) Sum of integrated intensities of hyperfine structure lines in range 93.17188-93.17613 GHz. \\
m) Integrated intensities are corrected for main-beam efficiency. \\
n) Profiles fitted with two Gaussians, but the total integrated intensity is not tabulated. \\
\end{flushleft}								
\end{deluxetable}
\end{landscape}



\newgeometry{bottom=1cm,left=1cm,right=1cm}

\begin{landscape}
\begin{deluxetable}{c c c c c c c}

\centering
\tablecolumns{7}
\tablewidth{460pt}
\tabletypesize{\footnotesize}

\tablecaption{\label{tbl:ColumnDensities}\sc Column Densities $^{a}$}

\tablehead{

	\multicolumn{1}{c}{Species (Transition)} &
	\multicolumn{1}{c}{$\nu$} &
	\multicolumn{1}{c}{$A_{ul}$} &
	\multicolumn{1}{c}{$T_{ex}$} &
	\multicolumn{1}{c}{Z(T)} &
	\multicolumn{1}{c}{$N_{X}$} &
	\multicolumn{1}{c}{$N_{X}/N_{^{13}CO}$} \\

	\multicolumn{1}{c}{} &
	\multicolumn{1}{c}{(GHz)} &
	\multicolumn{1}{c}{($s^{-1}$)} &
	\multicolumn{1}{c}{(K)} &
	\multicolumn{1}{c}{} &
	\multicolumn{1}{c}{(cm$^{-2}$)} &
	\multicolumn{1}{c}{} \\
}

\startdata

$^{13}$CO (2--1)$^{b}$ & 220.398686 & \ldots & 30 & \ldots & $6.98\times10^{14}$ & \ldots \\
                        & & & 20 & \ldots & $6.18\times10^{14}$ & \ldots \\
                        & & & 10 & \ldots & $7.28\times10^{14}$ & \ldots \\
                        & & & 6 & \ldots & $1.41\times10^{15}$ & \ldots \\
                        & & & 4 & \ldots & $4.65\times10^{15}$ & \ldots \\
C$^{18}$O (2--1) & 219.560357 & $6.012\times10^{-7}$ & 30 & 11.721 & $9.97\times10^{13}$ & \ldots \\
                        & & & 20 & 7.925 & $8.82\times10^{13}$ & \ldots \\
                        & & & 10 & 4.129 & $1.04\times10^{14}$ & \ldots \\
                        & & & 6 & 2.610 & $2.02\times10^{14}$ & \ldots \\
                        & & & 4 & 1.852 & $6.64\times10^{14}$ & \ldots \\
CS (5--4)        & 244.9356 & $2.981\times10^{-4}$ & 30 & 25.852 & $9.91\times10^{11}$ & $1.42\times10^{-3}$ \\
                        & & & 20 & 17.346 & $1.20\times10^{12}$ & $1.95\times10^{-3}$ \\
                        & & & 10 & 8.840 & $3.65\times10^{12}$ & $5.01\times10^{-3}$ \\
                        & & & 6 & 5.437& $2.49\times10^{13}$ & $1.76\times10^{-2}$ \\
                        & & & 4 & 3.736 & $3.92\times10^{14}$ & $8.43\times10^{-2}$ \\
HCN (3--2)       & 265.886431 & $8.356\times10^{-4}$ & 30 & 14.439 & $1.12\times10^{12}$ & $1.60\times10^{-3}$ \\
                        & & & 20 & 9.737 & $1.16\times10^{12}$ & $1.87\times10^{-3}$ \\
                        & & & 10 & 5.035 & $2.18\times10^{12}$ & $2.99\times10^{-3}$ \\
                        & & & 6 & 3.154 & $7.84\times10^{12}$ & $5.55\times10^{-3}$ \\
                        & & & 4 & 2.214 & $5.50\times10^{13}$ & $1.18\times10^{-2}$ \\
HCO$^{+}$ (3--2) & 267.557633 & $1.453\times10^{-3}$ & 30 & 14.351 & $9.12\times10^{11}$ & $1.31\times10^{-3}$ \\
                        & & & 20 & 9.678 & $9.47\times10^{11}$ & $1.53\times10^{-3}$ \\
                        & & & 10 & 5.006 & $1.80\times10^{12}$ & $2.47\times10^{-3}$ \\
                        & & & 6 & 3.137 & $6.54\times10^{12}$ & $4.63\times10^{-3}$ \\
                        & & & 4 & 2.202 & $4.65\times10^{13}$ & $9.99\times10^{-3}$ \\
DCN (3--2) & 217.2386 & $4.575\times10^{-4}$ & 30 & 17.598 & $<2.67\times10^{11}$ & $<3.82\times10^{-4}$ \\
                        & & & 20 & 11.843 & $<2.55\times10^{11}$ & $<4.14\times10^{-4}$ \\
                        & & & 10 & 6.088 & $<3.83\times10^{11}$ & $<5.25\times10^{-4}$ \\
                        & & & 6 & 3.786 & $<1.02\times10^{12}$ & $<7.24\times10^{-4}$ \\
                        & & & 4 & 2.635 & $<5.04\times10^{12}$ & $<1.09\times10^{-3}$ \\
DCO$^{+}$ (3--2) & 216.1126 & $7.658\times10^{-4}$ & 30 & 17.688 & $<1.23\times10^{11}$ & $<1.76\times10^{-4}$ \\
                        & & & 20 & 11.903 & $<1.17\times10^{11}$ & $<1.90\times10^{-4}$ \\
                        & & & 10 & 6.118 & $<1.75\times10^{11}$ & $<2.40\times10^{-4}$ \\
                        & & & 6 & 3.804 & $<4.64\times10^{11}$ & $<3.29\times10^{-4}$ \\
                        & & & 4 & 2.647 & $<2.27\times10^{12}$ & $<4.88\times10^{-4}$  \\
H$_{2}$CO (3--2) & 211.2115 & $2.271\times10^{-4}$ & 30 & 73.902 & $1.43\times10^{12}$ & $2.05\times10^{-3}$ \\
                        & & & 20 & 46.495 & $1.07\times10^{12}$ & $1.73\times10^{-3}$ \\
                        & & & 10 & 19.089 & $7.50\times10^{11}$ & $1.03\times10^{-3}$ \\
                        & & & 6 & 8.126 & $6.75\times10^{11}$ & $4.78\times10^{-4}$ \\
                        & & & 4 & 2.645 & $6.41\times10^{11}$ & $1.38\times10^{-4}$ \\
H$_{2}$CO (3--2) & 218.2222 & $2.818\times10^{-4}$ & 30 & 73.902 & $7.24\times10^{11}$ & $1.04\times10^{-3}$ \\
                        & & & 20 & 46.495 & $5.46\times10^{11}$ & $8.84\times10^{-4}$ \\
                        & & & 10 & 19.089 & $3.89\times10^{11}$ & $5.34\times10^{-4}$ \\
                        & & & 6 & 8.126 & $3.56\times10^{11}$ & $2.52\times10^{-4}$ \\
                        & & & 4 & 2.645 & $3.45\times10^{11}$ & $7.43\times10^{-5}$ \\
H$_{2}$CO (3--2) & 218.4756 & $1.571\times10^{-4}$ & 30 & 73.902 & $<9.27\times10^{11}$ & $<1.33\times10^{-3}$ \\
                        & & & 20 & 46.495 & $<6.99\times10^{11}$ & $<1.13\times10^{-3}$ \\
                        & & & 10 & 19.089 & $<4.98\times10^{11}$ & $<6.83\times10^{-4}$ \\
                        & & & 6 & 8.126 & $<4.56\times10^{11}$ & $<3.23\times10^{-4}$ \\
                        & & & 4 & 2.645 & $<4.43\times10^{11}$ & $<9.52\times10^{-5}$ \\
H$_{2}$CO (3--2) & 225.6978 & $2.772\times10^{-4}$ & 30 & 73.902 & $5.14\times10^{11}$ & $7.36\times10^{-4}$ \\
                        & & & 20 & 46.495 & $3.89\times10^{11}$ & $6.30\times10^{-4}$ \\
                        & & & 10 & 19.089 & $2.81\times10^{11}$ & $3.87\times10^{-4}$ \\
                        & & & 6 & 8.126 & $2.63\times10^{11}$ & $1.86\times10^{-4}$ \\
                        & & & 4 & 2.645 & $2.61\times10^{11}$ & $5.61\times10^{-5}$ \\

\enddata

\begin{flushleft}
a) Beam-averaged column densities assuming optically thin emission and adopting source radii (for all molecules) of $4''$ See \S~\ref{sec:CDE}. \\
b) $^{13}$CO column densities derived directly from the C$^{18}$O intensity assuming [$^{16}$O:$^{18}$O]/[$^{12}$C:$^{13}$C] = 7 \citep{2006A&A...456..675S,2005ApJ...634.1126M}.\\
\end{flushleft}
\end{deluxetable}
\end{landscape}

\restoregeometry



\begin{landscape}
\begin{deluxetable}{c c c c c c c c c}

\centering
\tablecolumns{9}
\tablewidth{560pt}
\tabletypesize{\footnotesize}

\tablecaption{\label{tbl:hfs}\sc Results of CN and C$_{2}$H Hyperfine Structure Analysis}

\tablehead{

	\multicolumn{1}{c}{Species} &
	\multicolumn{2}{c}{$v_{0}$ $^{e}$} &
	\multicolumn{1}{c}{$\Delta v$ $^{e}$} &
	\multicolumn{1}{c}{$\tau$ $^a$} &
	\multicolumn{1}{c}{$T_{\rm ex}$} &
	\multicolumn{1}{c}{$N_{\rm tot}/Q_{\rm tot}$ $^b$} &
	\multicolumn{1}{c}{$Q_{tot}(T_{\rm ex}$) $^c$} &
	\multicolumn{1}{c}{$N_{\rm tot}$ $^d$} \\

	\multicolumn{1}{c}{} &
	\multicolumn{1}{c}{(km s$^{-1}$)} &
	\multicolumn{1}{c}{(km s$^{-1}$)} &
	\multicolumn{1}{c}{(km s$^{-1}$)} &
	\multicolumn{1}{c}{} &
	\multicolumn{1}{c}{(K)} &
	\multicolumn{1}{c}{(cm$^{-2}$)} &
	\multicolumn{1}{c}{} &
	\multicolumn{1}{c}{(cm$^{-2}$)} \\
}

\startdata

    CN & $5.42\pm0.03$ & $7.38\pm0.03$ & $1.40\pm0.03$ & $2.23\pm0.45$ & $3.9^{+0.2}_{-0.2}$ & $4.9^{-0.3}_{+0.2}\times10^{12}$ & $10.97^{+0.42}_{-0.42}$ & $0.54^{-0.02}_{+0.01}\times10^{14}$ \\
    C$_2$H  & $5.18\pm0.11$ & $7.47\pm0.07$ & $1.39\pm0.06$ & $18.1\pm0.3$ & $2.9^{-0.1}_{+0.1}$ & $2.3^{-0.1}_{+0.0}\times10^{15}$ & $7.1^{+0.18}_{+0.18}$ & $1.6^{-0.31}_{+0.41}\times10^{16}$ \\

\enddata

\begin{flushleft}
a)  Total opacity obtained as the sum of individual hyperfine line opacities $\tau_{ul}$, where $\tau_{ul} = {\rm R.I.}\times \tau$.\\
b) From Equation~\ref{eq:HFSCDE}, \S~\ref{sec:CNC2H}.\\
c) From a linear interpolation to tabulated partition function values obtained from the CDMS database.\\
d) Column densities assuming a source diameter of $10''$ for both
molecules. See \S~\ref{sec:CNC2H}.\\
e) To make the two component Keplerian fit, we first fit one of the doppler-shifted components and then demanded that the intensity and the line width of the second component be the same as the first.\\
\end{flushleft}
\end{deluxetable}
\end{landscape}



\newgeometry{bottom=0.5cm,left=1cm,right=1cm}

\begin{deluxetable}{c c c c c c}

\centering
\tablecolumns{6}
\tablewidth{400pt}
\tabletypesize{\footnotesize}

\tablecaption{\label{tbl:ColumnDensityComparison}\sc Column Density Comparisons}

\tablehead{

	\multicolumn{1}{c}{} &
	\multicolumn{2}{c}{This Study} &
	\multicolumn{3}{c}{Previous Studies} \\

	\multicolumn{1}{c}{Species} &
	\multicolumn{1}{c}{$N_{X}$} &
	\multicolumn{1}{c}{$T_{ex}$} &
	\multicolumn{1}{c}{$N_{X}$} &
	\multicolumn{1}{c}{$T_{ex}$} &
	\multicolumn{1}{c}{Reference $^{a}$} \\

	\multicolumn{1}{c}{} &
	\multicolumn{1}{c}{(cm$^{-2}$)} &
	\multicolumn{1}{c}{(K)} &
	\multicolumn{1}{c}{(cm$^{-2}$)} &
	\multicolumn{1}{c}{(K)} &
	\multicolumn{1}{c}{} \\
}

\startdata

$^{13}$CO & $9.97\times10^{13}$ & 30 & $1.11\times10^{16}$ & 30 & 2 \\
                      & $6.18\times10^{14}$ & 20 & $3.6\times10^{14}$ & 25 & 3 \\
                      & $7.28\times10^{14}$ & 10 & $2.6\pm0.2\times10^{15}$ & 13.7 & 4 \\

C$^{18}$O & $9.97\times10^{13}$ & 30 & $3.31\times10^{15}$ & 30 & 2 \\

CS & $1.20\times10^{12}$ & 20 & $5.1\times10^{12}$ & 25 & 3 \\
      & $3.65\times10^{12}$ & 10 & $<1.2\times10^{13}$ & 10 & 6 \\
      & $2.49\times10^{13}$ & 6 & $8.7\pm1.6\times10^{12}$ & 8.7 & 7 \\

HCN & $1.16\times10^{12}$  & 20 & $1.8\times10^{12}$ & 25 & 3 \\
          & $2.18\times10^{12}$ & 10 & $4.3\times10^{12}$ & 10 & 6 \\
          &  $7.84\times10^{12}$ & 6 & $10.6\pm1.5\times10^{12}$ & $7.0\pm0.6$ & 8 \\
                       
HCO$^{+}$ & $9.12\times10^{11}$ & 30 & $9.25\times10^{12}$ & 30 & 2 \\
                       & $9.47\times10^{11}$ & 20 & $3.3\times10^{11}$ & 25 & 3 \\
                       & $9.47\times10^{11}$ & 20 & $8.0\pm0.5\times10^{12}$ & 19 & 4 \\
                       & $1.80\times10^{12}$ & 10 & $7.8\times10^{11}$ & 10 & 6 \\
                       
DCO$^{+}$ & $<1.17\times10^{11}$ & 20 & $<2.9\times10^{11}$ & 25 & 3 \\
                       & $<1.75\times10^{11}$ & 10 & $<1.1\times10^{11}$ & 10 & 6 \\
                       
H$_{2}$CO & $(0.39-1.07)\times10^{12}$ & 20 & $(0.7-1.9)\times10^{13}$ & 25 & 1 \\
                       & $(0.39-1.07)\times10^{12}$ & 20 & $7.2\times10^{12}$ & 20 & 3 \\
                      
CN & $(0.52-0.55)\times10^{14}$ & $3.9\pm0.2$ & $1.5\times10^{13}$ & 25 & 3 \\
       & $(0.52-0.55)\times10^{14}$ & $3.9\pm0.2$ & $3.4\times10^{13}$ & 10 & 6 \\
       & $(0.52-0.55)\times10^{14}$ & $3.9\pm0.2$ & $58\pm5\times10^{12}$ & $8.8\pm0.3$ & 8 \\

C$_{2}$H & $(1.29-2.01)\times10^{16}$ & $2.9\pm0.1$ & $2.9\pm1.1\times10^{13}$ & $6.3\pm1.4$ & 5 \\
                        
\enddata

\begin{flushleft}
a) References: 1. \citet{2003PASJ...55...11A};  2. \citet{2003ApJ...597..986Q}; 3. \citet{2004A&A...425..955T}; 4. \citet{2007A&A...467..163P}; 5. \citet{2010ApJ...714.1511H}; 6. \citet{2010A&A...524A..19F}; 7. \citet{2011A&A...535A.104D}; 8. \citet{2012A&A...537A..60C} \\
\end{flushleft}
\end{deluxetable}

\restoregeometry



\newgeometry{bottom=1cm}
\begin{landscape}
\begin{deluxetable}{c c c c c c c c c c}

\centering
\tablecolumns{10}
\tablewidth{580pt}
\tabletypesize{\footnotesize}

\tablecaption{\label{tbl:fracabun}\sc Comparison of Fractional Abundances Relative to $^{13}$CO, $N(X)/N(^{13}CO)$}

\tablehead{

	\multicolumn{1}{c}{Molecule} &
	\multicolumn{4}{c}{Protoplanetary Disk} &
          \multicolumn{4}{c}{YSO} &
          \multicolumn{1}{c}{Carbon Star} \\

	\multicolumn{1}{c}{} &
	\multicolumn{2}{c}{LkCa 15} & 
          \multicolumn{1}{c}{TW Hya} &
          \multicolumn{1}{c}{V4046 Sgr} &
          \multicolumn{1}{c}{IRAS 4A} &
          \multicolumn{1}{c}{IRAS 4B} &
          \multicolumn{1}{c}{IRS 7B} &
         \multicolumn{1}{c}{IRAS 16293} &
          \multicolumn{1}{c}{IRC +10216} \\

         \multicolumn{1}{c}{} &
	\multicolumn{1}{c}{This work $^{a}$} & 
          \multicolumn{1}{c}{1} &
          \multicolumn{1}{c}{2} &
          \multicolumn{1}{c}{2} &
          \multicolumn{1}{c}{3} &
          \multicolumn{1}{c}{3} &
          \multicolumn{1}{c}{4 $^{c}$} &
	\multicolumn{1}{c}{5 $^{c}$} &
          \multicolumn{1}{c}{6, 7} \\
}

\startdata

HCN & 1.87(-3) & 5.0(-3) & 2.6(-3) & 1.7(-3) & 1.4(-3) & 7.1(-4) & 5.6(-4) & 6.6(-4) & 0.56 \\
HCO$^{+}$ & 1.53(-3) & 9.17(-4) & \ldots & \ldots & 2.4(-4) & 3.3(-4) & 5.8(-4) & 8.4(-4) & 2.5(-3) $^{c}$ \\
H$_{2}$CO & 0.63-2.43(-3) & 0.39-4.72(-3) & \ldots & \ldots & 2.6(-3) & 3.1(-3) & 8.4(-4) & \ldots & \ldots \\
CS & 1.95(-3) & \ldots & 5.8(-3) & $<$7.0(-3) & 1.1(-3) & 7.9(-4) & 9.6(-4) & 1.8(-3) & 0.01 \\
CN & 0.01 & 0.04 & 0.40 & 0.08 & \ldots & \ldots & 9.6(-4) & 4.8(-5) & 0.01 \\ 
C$_{2}$H & 3.4 & \ldots & 21 & 8 & \ldots & \ldots & 3.2(-3) & 1.3(-4) & 0.09 \\ 

\enddata

\begin{flushleft}
a) abundances computed assuming $T_{ex}$ = 20 K, except for CN ($T_{ex} \approx$ 3.9 K) and C$_{2}$H ($T_{ex} \approx$ 2.9 K), where we adopt $N(^{13}CO)$ at  $T_{ex}$ = 4 K\\
b) 1. \citealt{2004A&A...425..955T}; 2. \citealt{2014ApJ...793...55K}; 3. \citealt{1995ApJ...441..689B}; 4.  \citealt{2012ApJ...745..126W}; 5. \citealt{2002A&A...390.1001S}; 6. \citealt{1994ApJS...94..147G}; 7. \citealt{2011ApJ...743...36P} \\
c) Values adapted from Table 5 of \citealt{2002A&A...390.1001S} (IRAS 16293), Table 4 of \citealt{2012ApJ...745..126W} (IRS 7B), and Table 5 of \citealt{2011ApJ...743...36P} (IRC +10216), assuming $N(H_{2})/N(^{13}CO)=6 \times 10^{5}$.
\end{flushleft}
\end{deluxetable}
\end{landscape}

\restoregeometry


\clearpage

\clearpage

\clearpage

\clearpage

\clearpage

\clearpage

\clearpage

\clearpage

\clearpage

\end{document}